\RequirePackage{fix-cm}

\documentclass[smallextended]{svjour3}       % onecolumn (second format)
\smartqed  % flush right qed marks, e.g. at end of proof
\def\preprintmode{1}

\usepackage{amsmath,amssymb,amsfonts}
\usepackage{algorithmic}
\usepackage{graphicx}
\usepackage{textcomp}
\usepackage{xcolor}
\def\BibTeX{{\rm B\kern-.05em{\sc i\kern-.025em b}\kern-.08em
    T\kern-.1667em\lower.7ex\hbox{E}\kern-.125emX}}

\usepackage{scalerel}
\usepackage{tikz}
\usetikzlibrary{svg.path}

\usepackage[hidelinks]{hyperref}

\usepackage[utf8]{inputenc}
\usepackage[T1]{fontenc}
\usepackage[authoryear,sort&compress]{natbib}
\usepackage{color, colortbl}
\usepackage{arydshln}
\usepackage{tabularx}
\newcolumntype{Y}{>{\centering\arraybackslash}X}
\usepackage{subcaption}
\usepackage{ifthen}

\usepackage[switch]{lineno}
\usepackage[many]{tcolorbox}    	% for COLORED BOXES (tikz and xcolor included)

\usepackage{pdflscape}
\usepackage{adjustbox}
\usepackage{afterpage}
\usepackage{enumitem}
\usepackage[normalem]{ulem}

\newcommand{\qt}[1]{\textit{``#1''}}
\newcommand{\myquestion}[2]{\begin{enumerate}[leftmargin=1cm]\item[\textbf{#1}]\textit{#2}\end{enumerate}}
\newcommand{\rqn}[1]{RQ\textsubscript{#1}}

\newcommand{\chline}[2]{\cline{#1}\multicolumn{#2}{c}{}\\[-6pt]\hline}

\newcommand{\hhline}{\hline\hline}

\newcommand{\citesingpossesive}[1]{\citeauthor{#1}'~(\citeyear{#1})}

\renewcommand{\email}[1]{\href{mailto:#1}{#1}}

\newcommand{\ifEmpty}[2]{%
    \if\relax\detokenize{#2}\relax% Check if #1 is empty
        {#1}
    \else
        {#2}
    \fi
}

\newcommand{\revOneAdd}[1]{\ifEmpty{}{{\color{blue}{#1}}}}
\newcommand{\revOneRem}[1]{\ifEmpty{}{{\color{red}\sout{#1}}}}

\renewcommand{\revOneAdd}[1]{#1}
\renewcommand{\revOneRem}[1]{}

\definecolor{Gray}{gray}{0.9}
\definecolor{LightGray}{gray}{0.92}
\definecolor{LighterGray}{gray}{0.94}

\tcbset{
    sharp corners,
    colback = white,
    before skip = 0.2cm,    % add extra space before the box
    after skip = 0.5cm      % add extra space after the box
}       

\definecolor{ggplot_blue}{rgb}{0.203, 0.541, 0.741}
\definecolor{ggplot_fg}{rgb}{0.957, 0.976, 0.988}
\newtcolorbox{takeaway}{
    colback=gray!10!white,
    boxsep=2pt,left=1pt,right=1pt,top=2pt,bottom=2pt,
    arc=100pt}

\newenvironment{insight}[1]{\begin{takeaway}\textbf{Answering~\rqn{#1}:}}{\end{takeaway}}

\definecolor{orcidlogocol}{HTML}{A6CE39}
\tikzset{
  orcidlogo/.pic={
    \fill[orcidlogocol] svg{M256,128c0,70.7-57.3,128-128,128C57.3,256,0,198.7,0,128C0,57.3,57.3,0,128,0C198.7,0,256,57.3,256,128z};
    \fill[white] svg{M86.3,186.2H70.9V79.1h15.4v48.4V186.2z}
                 svg{M108.9,79.1h41.6c39.6,0,57,28.3,57,53.6c0,27.5-21.5,53.6-56.8,53.6h-41.8V79.1z M124.3,172.4h24.5c34.9,0,42.9-26.5,42.9-39.7c0-21.5-13.7-39.7-43.7-39.7h-23.7V172.4z}
                 svg{M88.7,56.8c0,5.5-4.5,10.1-10.1,10.1c-5.6,0-10.1-4.6-10.1-10.1c0-5.6,4.5-10.1,10.1-10.1C84.2,46.7,88.7,51.3,88.7,56.8z};
  }
}
\newcommand{\orcidlink}[1]{\href{https://orcid.org/#1}{#1}}
\newcommand\orcidicon[1]{\href{https://orcid.org/#1}{\mbox{\scalerel*{
\begin{tikzpicture}[yscale=-1,transform shape]
\pic{orcidlogo};
\end{tikzpicture}
}{|}}}}

\begin{document}

\if 0\preprintmode
    \journalname{Empirical Software Engineering}
\else
    \journalname{Preprint}
\fi

\newcommand{\grantref}[0]{\thanks{This manuscript is a revision of the first author's master's thesis~\citep{meijer_influence_2023}, supervised and graded by the second and third authors. This work was partially supported by the Wallenberg AI, Autonomous Systems and Software Program (WASP) funded by the Knut and Alice Wallenberg Foundation.}}

\title{Ecosystem-wide influences on pull request decisions: insights from NPM\grantref}
% \subtitle{insights from NPM}
% \titlerunning{Ecosystem-wide Influences on Pull Request Decisions: Insights from NPM}

\author{Willem Meijer \and Mirela Riveni \and Ayushi Rastogi}
\authorrunning{Willem Meijer, Mirela Riveni and Ayushi Rastogi} % if too long for running head
% Regarding adding authors during the review process; taken from the EMSE submission guidelines: Authors are strongly advised to ensure the author group, the Corresponding Author, and the order of authors are all correct at submission. Adding and/or deleting authors during the revision stages is generally not permitted, but in some cases may be warranted. Reasons for changes in authorship should be explained in detail. Please note that changes to authorship cannot be made after acceptance of a manuscript.

\newcommand{\authorEntry}[5]{#1~\at~#2\\#3\\E-mail:~\email{#4}\\\orcidicon{#5}~\orcidlink{#5}\and}
% Don't put extra newlines into this; it'll break compilation.
\institute{\authorEntry{Willem Meijer}{Corresponding author\\University of Groningen / Linköping University}{Groningen, The Netherlands / Linköping, Sweden}{willem.meijer@liu.se}{0000-0001-8482-3917}
            \authorEntry{Mirela Riveni}{University of Groningen}{Groningen, The Netherlands}{m.riveni@rug.nl}{0000-0002-4991-3455}
            % \authorEntry{D{\'a}niel Varr{\'o}}{Linköping University}{Linköping, Sweden}{daniel.varro@liu.se}{0000-0002-8790-252X}
            \authorEntry{Ayushi Rastogi}{University of Groningen}{Groningen, The Netherlands}{a.rastogi@rug.nl}{0000-0002-0939-6887}
}

\if 0\preprintmode
    \date{Received: date / Accepted: date}
\else
    \date{\textit{\today.}}
\fi

\maketitle

    \begin{abstract}
        The pull-based development model facilitates global collaboration within open-source software projects. 
\revOneAdd{However, whereas it is increasingly common for software to depend on other projects in their ecosystem,} most research on the pull request decision-making process explored factors within projects, not the broader software ecosystem they comprise.
We uncover ecosystem-wide factors that influence pull request acceptance decisions.
We collected a dataset of approximately $1.8$ million pull requests and $2.1$ million issues from 20,052 GitHub projects within the NPM ecosystem.
Of these, $98\%$ depend on another project in the dataset, enabling studying collaboration across dependent projects.
We employed social network analysis to create a collaboration network in the ecosystem, and mixed effects logistic regression and random forest techniques to measure the impact and predictive strength of the tested features.
We find that gaining experience within the software ecosystem through active participation in issue-tracking systems, submitting pull requests, and collaborating with pull request integrators and experienced developers benefits all open-source contributors, especially project newcomers.
\revOneAdd{These results are complemented with an exploratory qualitative analysis of $538$ pull requests.
We find that developers with ecosystem experience make different contributions than users without.
For example, they introduce new features and bug fixes less commonly than dependency updates as part of maintenance. 
Zooming in on a subset of $111$ pull requests with clear ecosystem involvement, we find 3 overarching and 10 specific reasons why developers involve ecosystem projects in their pull requests.
For example, when another project has implemented a solution that can be used as a reference implementation.}
The results show that combining ecosystem-wide factors with features studied in previous work to predict the outcome of pull requests reached an overall F1 score of $0.92$.
\revOneAdd{However, the outcomes of pull requests submitted by newcomers are harder to predict.}

        % You can only have 4 - 6 index terms.
        \keywords{open-source software
            \and software ecosystem
            \and social coding platform
            \and software package
            \and software dependency
            \and project newcomer
            \and collaborative software engineering
            % \and mining software repositories
            % \and social network analysis
        }
    \end{abstract}
    
    \section{Introduction}
\label{sec:introduction}

\revOneAdd{
The increasing interconnectivity between open-source software \citep{decan_empirical_2019} has made studying collaboration across multiple projects paramount.
Where we could once rely on the integrity of a project itself, modern open-source software increasingly depends on functionality built by others \citep{decan_empirical_2019}.
This means that problems introduced in one project will inevitably propagate, potentially causing a cascade of issues for downstream developers and users \citep{ma_impact_2020}.
This would not be a problem if developers could account for this trivially.
However, as the complexity of software continues to increase, so does the skill required to create and maintain them, and consequently, the ability and required capacity to account for each others’ mistakes and help each other resolve them.}

This article studies open-source software ecosystems, \qt{a network of open-source software communities working on a common technology} \citep{franco-bedoya_open_2017, hanssen_theoretical_2012}.
The NPM package library is an example of this, containing millions of JavaScript projects, many of which reuse each others’ functionalities.
This common technology allows developers to share knowledge across projects; something that has in the past been reported to positively affect project downloads \citep{mendez-duron_returns_2009, fershtman_direct_2011, peng_network_2013} and longevity \citep{wang_survival_2012, qiu_going_2019, valiev_ecosystem-level_2018, wattanakriengkrai_giving_2023}.
However, even though we know many things about collaborating within individual projects \citep{gharehyazie_social_2013, gharehyazie_developer_2015, carillo_what_2017, bosu_impact_2014, qiu_going_2019, zoller_topology_2020}, only limited research has been conducted to understand what collaboration looks like at an ecosystem scale and how we can improve this collaboration.
This paper, then, attempts to capture parts of ecosystem-wide collaboration dynamics, specifically focusing on the impact of cross-project collaboration on pull request decisions.

The pull-based development model \citep{gousios_exploratory_2014, gousios_work_2015} is a frequently used development model that introduced the notion of pull requests.
It allows developers to separate development from decision-making such that developers create pull requests to propose changes and project maintainers can choose to accept/reject them.
Previous research has significantly contributed to our understanding of pull request decisions \citep{zhang_pull_2022-1}, addressing factors of pull requests themselves, like the number of commits a pull request has \citep{gousios_exploratory_2014, yu_determinants_2016, soares_acceptance_2015, soares_rejection_2015, khadke_predicting_2012, kononenko_studying_2018, zampetti_study_2019}, or intra-project factors (i.e., factors measured within a project), like the number of previously accepted pull requests of a developer \citep{gousios_exploratory_2014, rastogi_relationship_2018, khadke_predicting_2012, zampetti_study_2019, zhang_pull_2022}.
Notably, only one previous study considered the ecosystem perspective \revOneAdd{in this context} \citep{dey_effect_2020}, identifying an early positive relationship between ecosystem participation from a technical standpoint and pull request acceptance, thus laying the groundwork for further study.

\revOneAdd{
Our study first aims to verify the results by \citet{dey_effect_2020} through a coarse- and fine-grained analysis of the number of developers' ecosystem contributions measured using coding and non-coding contributions.
We do this by first studying \textit{general} ecosystem contributions which we then subdivide into \textit{upstream}, \textit{downstream}, and \textit{non-dependency} contributions.
This provides valuable insights into the impact of an individual's ecosystem resume by counting the \textit{number} of things they did and \textit{where} they did them.
However, because such metrics restrict themselves to the work of individuals, they provide limited insights into greater \textit{collaborative} efforts in the community.
To capture this, we study the impact of \textit{direct collaboration} between contributors in the ecosystem and the \textit{ecosystem-wide community standing} of contributors.
Respectively, these highlight joint efforts on a smaller and larger scale.
Using these metrics, we can extend our research by studying how contributors move through software ecosystems and start new collaborations, allowing us to do an early analysis of the impacts of an individual's \textit{technical resume} and \textit{past collaborations} in the ecosystem on joining new projects.
Specifically, we identify the impact of ecosystem participation on the pull request submissions done by \textit{newcomers}.
Consequently, this paper lays the groundwork for understanding the collaboration dynamics in open-source software ecosystems that can be leveraged to benefit projects and to-be contributors.
}

\section{Research Questions}

% RQ ecosystem experience
Our work is inspired by and builds on the work of \citet{dey_effect_2020}, who suggested that the experience acquired by making code-related contributions in an open-source software ecosystem increases the chance of pull request acceptance in another project in the ecosystem.
Although software communities are commonly defined as a group of coders, various studies have shown the prominence of non-coding collaborators \citep{trinkenreich_hidden_2020, canovas_izquierdo_analysis_2021, geiger_labor_2021, trinkenreich_pots_2022}, like bug reporters or community managers, who participate in issue-tracker discussions \citep{panichella_how_2014}.
Intuition suggests that non-coding contributions positively relate to decision outcomes, as they are used to discuss domain-relevant information, like architectural knowledge \citep{soliman_exploratory_2021} or requirements \citep{perez-verdejo_requirements_2021}, and often complement information stored in pull requests \citep{alshara_pi-link_2023}.
However, no formal study has been done to test the impact of non-coding contributions on pull request decisions.
\revOneAdd{This highlights a group of contributors who are a key component of collaboration in open-source communities but were overlooked in previous related studies.}
Therefore, beyond replicating \citesingpossesive{dey_effect_2020} findings, we include non-coding contributions to verify this intuition.
Thus, our first research question is: \revOneAdd{\myquestion{\rqn{1}}{How do a developer's \textbf{coding and non-coding contributions in the ecosystem} affect the acceptance of their pull request submissions?}}

% RQ dependency experience
A unique aspect of software ecosystems is that projects have technical dependencies, such that a project reuses the functionalities contained in another \citep{decan_empirical_2019}.
An example of this is a web application built using the React framework, reusing the framework's functionality and extending it wherever necessary.
\revOneAdd{In other words, the project depends on React.
For React, the web application is a downstream project, while React is an upstream project of the web application.

Previous research has indicated the benefit of collaboration between upstream and downstream projects because the knowledge required to implement a package could help to improve and extend it \citep{valiev_ecosystem-level_2018, maeprasart_understanding_2023, palyart_study_2018, rehman_newcomer_2022, shah_motivation_2006, subramanian_analyzing_2022}, and experience building a package could help implement it \citep{bogart_how_2016, bogart_when_2021}.
\revOneAdd{Further, congruence between developer groups and dependency networks} has been shown to positively affect project longevity \citep{valiev_ecosystem-level_2018, wattanakriengkrai_giving_2023}.

Although the coarse-grained analysis for \rqn{1} indicates the impact of having generally related experience, we believe specialized knowledge acquired by actively working with upstream/downstream projects could have additional benefits.
In other words, although gaining experience in the ecosystem has an impact, we argue that \textit{where} you acquired that experience matters too.}

The work of \citet{dey_effect_2020} claimed that \qt{if any of the projects the pull request creator previously contributed to depend on the repository to which the pull request is being created, it is more likely to be accepted} \revOneAdd{(i.e., making code contributions to downstream projects).
Although this gives us an expectation about the impact of downstream contributions, the same has not been studied for upstream projects, motivating us to include it in our study.}
To contextualize upstream and downstream contributions, we include non-dependency contributions as well.
Non-dependent projects are projects the pull request creator previously contributed to that do not have an upstream or downstream dependency on the project where the pull request is submitted.
Including non-dependent projects is crucial because it enables us to assess the additional benefits gained from experience in closely related projects.
Studying this is a direct continuation of \rqn{1}, providing fine-grained insights into the added benefits of technical dependencies by zooming in on \revOneAdd{where developers acquired their experience.}
Therefore, our second research question is: \revOneAdd{\myquestion{\rqn{2}}{How do a developer's contributions to \textbf{upstream and downstream projects} affect the acceptance of their pull request submissions?}}

% RQ collaboration.
\revOneAdd{The first two research questions give us a clear understanding of the impact of \textit{whether} and \textit{where} developers have previously contributed to the ecosystem, giving a clear overview of a developer's resume.
However, this limitedly explains the impact of \textit{collaboration} in the ecosystem because it is restricted to the work of individuals.}
Prior studies have shown that the intra-project social connectedness can positively affect the pull request acceptance \citep{gharehyazie_social_2013, gharehyazie_developer_2015, carillo_what_2017, bosu_impact_2014}, project longevity \citep{qiu_going_2019}, general productivity \citep{zoller_topology_2020}, at the risk of increased bug rates \citep{chen_empirical_2024}.
Similarly, the \qt{social distance} between pull request submitters and their integrators, measured by whether the submitter followed the integrator before submitting the pull request, influences the decision outcome positively \citep{zhang_pull_2022, yu_determinants_2016, tsay_influence_2014, iyer_effects_2021, soto_analyzing_2017}.
Although these studies explain the importance of social connections within projects, and to some extent outside projects, they give no insights into the relevance of connections \revOneAdd{created through collaboration.}

We argue that these connections could also influence pull request decision-making at the ecosystem level.
Compared to the follower network \citep{zhang_pull_2022, yu_determinants_2016, tsay_influence_2014, iyer_effects_2021, soto_analyzing_2017, celinska_coding_2018}, we include two stronger indicators of professional relationships: \textit{direct collaboration} and \textit{ecosystem-wide community standing}.
This is measured by constructing a social network using developers' development activities, enabling us to study the impact of collaboration at an ecosystem scale.
Our third research question is: \revOneAdd{\myquestion{\rqn{3}}{How do a developer's \textbf{ecosystem-wide community standing and direct collaboration with their integrator in the ecosystem} affect the acceptance of their pull request submissions?}}

% RQ newcomers
Finally, studying software ecosystems yields the opportunity to study \qt{experienced project newcomers;} developers who are new to a project but have previously made contributions inside the wider ecosystem.
\revOneAdd{Including newcomers enables us to study how contributors move through the ecosystem, as they join a project for the first time, starting new collaborations.}
Newcomers are paramount to open-source development as joining and leaving a project is easy \citep{forte_defining_2013} and they can extend project longevity \citep{qiu_going_2019}.
Regardless, changes proposed by newcomers are more commonly refused \citep{kovalenko_code_2018, lee_are_2017, soares_rejection_2015, soares_acceptance_2015}, and large projects can be reluctant to accept newcomers \citep{palyart_study_2018}.
Various reasons have been identified as to why newcomers decide to discontinue contributing even though they were motivated to contribute at the start \citep{steinmacher_social_2015, steinmacher_almost_2018, steinmacher_let_2019, steinmacher_overcoming_2019, wermke_committed_2022, rehman_newcomer_2022, rastogi_ramp-up_2015, geiger_labor_2021}, like a lack of knowledge or poor communication behaviors.
This leaves the question of whether newcomers who \textit{have had} the opportunity to hone these skills have a different onboarding process.

Although intra-project socialization/collaboration behavior was shown to improve onboarding \citep{gharehyazie_social_2013, gharehyazie_developer_2015, carillo_what_2017} and that people tend to join projects and ecosystems they are socially connected to \citep{casalnuovo_developer_2015, hahn_emergence_2008, de_souza_social_2016}, no study has been conducted on the effect of ecosystem-wide socialization on pull request acceptance decisions submitted by newcomers.
\citet{rastogi_how_2021} hypothesized that changes proposed by experienced newcomers are more likely to be accepted, however, do not provide any empirical proof for this.
Therefore, we address \rqn{1--3} specifically for newcomers, operationalizing \citeauthor{rastogi_how_2021}' hypothesis by asking: \revOneAdd{\myquestion{\rqn{4}}{How do \textbf{a project newcomer's contributions and collaborations in the ecosystem} affect the acceptance of their pull request submissions?}}

\section{Related Work}
\label{sec:related-work}

\revOneAdd{
    Automating part of pull request reviews could improve efficiency in software projects, making pull request decision prediction \citep{zhang_pull_2022}, together with the timeliness of this decision \citep{zhang_pull_2022-1}, one of the main challenges in modern code reviews \citep{badampudi_modern_2023}.
    Although the review itself is important, the ability of submitters to respond to the review is equally relevant.
    Various studies have addressed the impact of unresponsiveness or inappropriate responsiveness on pull request outcomes, indicating its relationship to pull request abandonment and staleness \citep{li_opportunities_2022} and whether the submitter continues participating afterward \citep{steinmacher_social_2015, steinmacher_almost_2018, steinmacher_let_2019}.
    Although some of these challenges have existed for some time already, new challenges and opportunities have appeared as well.
    For example, various studies have been conducted on large language models in software engineering \citep{hou_large_2024}, which are used to generate pull request reviews \citep{lu_llama-reviewer_2023} and pull request descriptions \citep{xiao_generative_2024}.
    Our work extends this extensive body of literature by assessing how ecosystem-wide participation affects pull request decisions.
}

\newcommand{\featureCommits}{PR commit count}
\newcommand{\featureAge}{PR age in minutes}
\newcommand{\featureIntegrator}{PR integrator experience}
\newcommand{\featureSelfIntegrated}{PR is self-integrated}
\newcommand{\featureComments}{PR has comments}
\newcommand{\featureExternalComments}{\revOneAdd{PR has comment from external contributor}}
\newcommand{\featureReference}{\revOneAdd{PR description contains \qt{\#}}}
\newcommand{\featureNewcomer}{\revOneAdd{PR submitter is a newcomer}}

{

\newcolumntype{K}{>{\hsize=0.18\hsize}X}
\newcolumntype{V}{>{\hsize=0.82\hsize}X}

\begin{table}[!t]
    \centering
    \small
    \caption{Overview of control variables for \revOneAdd{pull request (PR) decisions} suggested by \citet{zhang_pull_2022} that are used in this study}
    \begin{tabularx}{\columnwidth}{|K|V|}
        \hline
        \rowcolor{Gray}
        \textbf{Name} & \textbf{Description} \\
        \hhline
        \textit{\featureCommits} & 
            The number of commits in a PR \citep{gousios_exploratory_2014, yu_determinants_2016, soares_acceptance_2015, soares_rejection_2015, khadke_predicting_2012, kononenko_studying_2018, zampetti_study_2019}.
            This has differing impacts in the literature. \\

        \rowcolor{LighterGray}
        \textit{\featureAge} & 
            The time between the PR was opened and closed \citep{zampetti_study_2019, soares_acceptance_2015, legay_impact_2019}.
            \revOneAdd{Older PRs are more likely to be rejected.} \\
        
        \textit{\featureIntegrator} & 
            The number of PRs the integrator reviewed in the project \citep{baysal_investigating_2016}.
            \revOneAdd{More experienced integrators are more likely to accept a PR.} \\
        
        \rowcolor{LighterGray}
        \textit{\featureSelfIntegrated} & 
            Whether the PR is submitted and integrated by the same person \citep{zhang_pull_2022}.
            \revOneAdd{PRs are more likely to be rejected when the submitter and integrator are the same person.} \\

        \textit{\featureComments} & 
            Whether the PR has comments \citep{soares_rejection_2015, golzadeh_effect_2019}.
            \revOneAdd{PRs with comments are more likely to be rejected.} \\

        \rowcolor{LighterGray}
        \textit{\featureExternalComments} &
            Whether someone different than the submitter, integrator or a project contributor \revOneAdd{(i.e., an external contributor)} commented on the PR \citep{golzadeh_effect_2019}.
            \revOneAdd{PRs with comments from an external are more likely to be accepted.} \\
        
        \textit{\featureReference} & 
            Whether the PR's title or description references another issue or pull request \citep{gousios_exploratory_2014, yu_determinants_2016}. \revOneAdd{In GitHub, this is done using a \qt{\#}.} \revOneAdd{When a PR contains a reference, it is more likely to be accepted.} \\

        \hdashline

        \rowcolor{LighterGray}
        \textit{\featureNewcomer} & 
            Whether the PR submitter has successfully \revOneAdd{committed to the project in the past} \citep{soares_acceptance_2015, soares_rejection_2015, lee_are_2017, kovalenko_code_2018}.
            \revOneAdd{PRs submitted by newcomers are more likely to be rejected.}
            This feature was not part of the principal features reported by \citet{zhang_pull_2022} but was added as it motivates \rqn{4}. \\
        
        \hline
    \end{tabularx}
    \label{tab:control-variables}
\end{table}

}

Our work is motivated by a recent literature study \citep{zhang_pull_2022} identifying 94 factors influencing pull request decisions.
Most of these are intra-project factors (i.e., factors measured inside a project) like the number of pull requests that someone submitted in the project \citep{gousios_exploratory_2014, khadke_predicting_2012, rastogi_relationship_2018, zampetti_study_2019, zhang_pull_2022}.
\citeauthor{zhang_pull_2022} ranked these factors based on their importance and found that they changed based on context.
For example, the pull request integrator's experience is less important for self-integrated pull requests.
Table~\ref{tab:control-variables} overviews the variables included in this study, which are the principal factors reported by \citet{zhang_pull_2022} for general pull request decision research.
Although many decision factors are technical, relevant human factors were found as well, like interacting with core developers \citep{yu_determinants_2016}, geographical location \citep{rastogi_relationship_2018, rastogi_biases_2016} and gender \citep{terrell_gender_2017}.

\citet{gharehyazie_social_2013, gharehyazie_developer_2015} studied the impact of intra-project socialization in mailing lists in Apache projects.
They find that socialization --- in particular two-way communication --- is an important positive indicator of whether a contributor will become a \qt{committer} (a type of membership in Apache projects).
\revOneAdd{
    Further, \citet{bosu_impact_2014} studied how a developer's community standing within a project affects different aspects of their code reviews.
    They calculated community standing using various social network analysis metrics using a social network based on review interactions.
    They find that core developers have to wait significantly less long to receive their first feedback, the review duration is shorter, and their proposals are more commonly accepted.
}

Some studies explored the effect of project-transcending factors \citep{tsay_influence_2014, yu_determinants_2016, iyer_effects_2021, soto_analyzing_2017, zhang_pull_2022, celinska_coding_2018}.
Most of these \citep{tsay_influence_2014, iyer_effects_2021, yu_determinants_2016, soto_analyzing_2017, zhang_pull_2022} evaluated GitHub's follower feature, taking this as a proxy for social connectedness on the platform.
They found that being followed by others and following others \citep{soto_analyzing_2017, yu_determinants_2016,zhang_pull_2022}, and following your integrator \citep{tsay_influence_2014, iyer_effects_2021,zhang_pull_2022} positively impact pull request acceptance.
Finally, \citet{celinska_coding_2018} found a positive relationship between follower and collaboration networks on the number of pull requests and forks a project receives.

The works of \citet{dey_effect_2020}, \citet{cheng_developer_2017}, and \citet{jergensen_onion_2011}, and lie closest to our work.
\citeauthor{dey_effect_2020} studied the effect of ecosystem experience on pull request acceptance, using three ecosystem factors: the submitter's pull request track record (pull request count and acceptance rate), their general experience (the number of commits and projects worked on), and whether they worked on a downstream project.
They showed that all factors, except pull request count, positively affect pull request acceptance.
However, this study does not account for the difference between intra-project and ecosystem-wide experience, as these metrics were aggregated in their analysis.
\revOneAdd{Because previous work suggested the impact of intra-project experience \citep{gousios_exploratory_2014, khadke_predicting_2012, rastogi_relationship_2018, zampetti_study_2019, zhang_pull_2022}, these results are likely confounded.}
% Because previous work suggested the impact of intra-project experience \citep{gousios_exploratory_2014, khadke_predicting_2012, rastogi_relationship_2018, zampetti_study_2019, zhang_pull_2022}, thus confounding the results.

\citet{cheng_developer_2017} and \citet{jergensen_onion_2011} both studied the impact of ecosystem experience on becoming a member of GNOME projects.
\citet{cheng_developer_2017} showed that having more contributions and projects worked on positively affects becoming a core member, indicated by the number of commits they make and the number of people they collaborate with inside a project.
In contrast, \citet{jergensen_onion_2011} showed that developers with ecosystem experience, indicated by the number of releases they were active on in other projects, do not change a system's core functionality more often, which they argue is preserved for core project members, whereas developers with intra-project experience do.
Although these studies highlight the impact of ecosystem-wide contributions on the types and number of code changes, these findings cannot be translated to the pull-based development model.

We further differentiate ourselves by addressing coding and non-coding contributions, upstream and downstream dependencies, and ecosystem-wide collaborations.
We specifically test their impact on pull requests submitted by project newcomers.
    
    \section{Methodology}
\label{sec:methodology}

\begin{figure}[b]
    \centering
    \includegraphics[width=0.7\columnwidth]{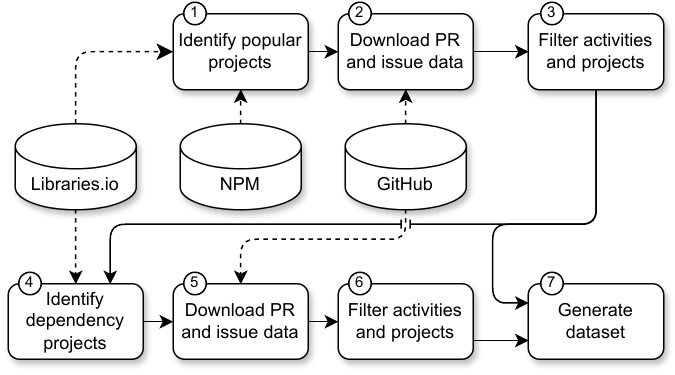}
    \caption{Visualization of the data collection process.
    For more details, refer to Figure 1 in the original master's thesis \citep{meijer_influence_2023}
    }
    \label{fig:data-collection-process}
\end{figure}

\revOneAdd{
    The following sections describe this article's methodology.
    We first introduce the data collection process in Section~\ref{sec:data-collection}, followed by this study's metrics in Section~\ref{sec:variable-construction}.
    Finally, Section~\ref{sec:data-analysis} describes the quantitative evaluation of these metrics, and Section~\ref{sec:quaitative-analysis} the exploratory qualitative study.
}

\subsection{Data Collection}
\label{sec:data-collection}

    Open-source software projects commonly use a combination of pull requests and issue-tracking systems \citep{alshara_pi-link_2023} to make coding and non-coding contributions, respectively.
    Consequently, three data sources are required to answer the research questions: pull requests, issues, and a list of project dependencies from a software ecosystem.
    Although various projects use media like mailing lists \citep{panichella_how_2014}, these are excluded as they are not integrated into platforms like GitHub.
    \revOneAdd{Similarly, we do not include code commits \citep{dey_effect_2020} to limit the scope of this study.
    However, code commits and mailing lists are likely fruitful considerations for future work.}
    Issues have been included because non-coding contributions are prominent in open-source communities \citep{trinkenreich_hidden_2020, canovas_izquierdo_analysis_2021} but have not been formally studied yet.
    The dataset created by \citet{katz_librariesio_2020} is the starting point of this research, containing project archive data of 32 software package ecosystems (like NPM, Maven, and Go).
    It contains information like package names, repository descriptions, and dependencies.
    Although \citesingpossesive{katz_librariesio_2020} dataset only contains data up to 2020, it is valuable as creating a similar dataset is very time-consuming.
    We zoom into the NPM ecosystem, a JavaScript package library.
    We chose NPM for three reasons: 1)~\citet{dey_effect_2020} used it as well, meaning our results can be compared, 2)~it is the largest ecosystem in the dataset, and 3)~NPM packages have explicit dependencies on each other.
    Since this dataset does not contain pull requests and issues, we collected these separately using GrimoireLab \citep{duenas_grimoirelab_2021}.
    Figure~\ref{fig:data-collection-process} visualizes the data collection process elaborated in the following sections.

    \subsubsection{Project Sampling}
    \label{sec:project-selection}

        \citesingpossesive{katz_librariesio_2020} dataset contains archival data of $1.2$ million NPM packages.
        Because this is too many to process completely, we sampled their dataset.
        This is done in two stages: collecting development activities from 1)~popular projects and 2)~ upstream/downstream projects of the popular projects.
        Respectively, these ensure that our dataset contains many development activities and many projects involved in a dependency.
        To simplify this process, a simple set of inclusion criteria was defined: 1)~projects must have a GitHub repository, 2)~they cannot be a fork of another project, and 3)~the project must have at least five \qt{valid} pull requests to ensure that the project systematically uses pull requests (elaborated in Section~\ref{sec:activity-selection}).
        
        Popular projects were sampled using one additional criterion: \qt{the project must have at least 10,000 downloads in the past $16$ months.} 
        This definition aligns with \citet{dey_effect_2020} and is a proxy for project activity, assuming that popular projects have more development activities.
        \revOneAdd{Similarly, upstream/downstream projects were sampled with one additional criterion: \qt{the project must be an upstream or downstream project of a popular project.}
        Although the number of upstream projects was sufficiently small to collect completely (approximately 12,500 projects), the number of downstream projects was too large (approximately 445,000 projects).
        Therefore, we randomly sampled downstream projects.
        A preliminary analysis showed that downstream projects have approximately half the number of pull requests compared to upstream projects.
        Therefore, the sample of downstream projects was approximately doubled to avoid bias towards upstream projects.
        Popular projects and upstream projects were \textit{not} subsampled.
        Following the guidelines for random sampling described by \citet{lohr_sampling_2021} with a margin of error of $5\%$ and a confidence interval of $95\%$, a minimum of $950$ projects should be sampled.
        Therefore, because our sample size is about 12,500, our sample is representative of downstream projects.}

    \subsubsection{Activity Filtering}
    \label{sec:activity-selection}

        \revOneAdd{To ensure the quality of our dataset,} development activities (i.e., pull requests and issues) were removed using the following criteria:
        \begin{itemize}
            \item
                \textit{They cannot be submitted by a deleted account.}
                When users delete their accounts, their development activities are assigned to a \qt{ghost} user, who has many activities assigned to it.

            \item
                \textit{The development activity is closed.}
                This ensures that the discussion in the activity has come to fruition.

            \item
                \textit{The development activity has no missing data.}
                Some development activities miss data like user details.

            \item
                \textit{They cannot be submitted by a bot.}
                We removed bots because \citet{dey_effect_2020} attributed some of the quirks in their results to bots.
        \end{itemize}

    \subsubsection{Bot Removal}
    \label{sec:bot-removal}
    
        Many actions in GitHub can be automated using bots \citep{rombaut_theres_2023}.
        An example is \qt{Dependabot} which can be used to manage dependency updates.
        When unaccounted for, bots can affect results in developer-oriented research \citep{dey_effect_2020}.
        \revOneAdd{However, identifying} bots is not a trivial task.
        GitHub metadata was used as a first line of defense, as creators can mark their accounts as bots.
        However, this method does not work perfectly because various bots use user accounts.
        Although bot detection is an active field study \citep{dey_detecting_2020, golzadeh_ground-truth_2021, golzadeh_bot_2020, chidambaram_bot_2022}, none of the proposed classifiers could be directly applied to this study due to stringent data requirements and the substantial amount of required manual effort.
        Fortunately, the classifications of \citet{dey_detecting_2020, dey_dataset_2020} and \citet{golzadeh_ground-truth_2021, golzadeh_ground-truth_2020} are publicly available, and all bots contained in their datasets were removed.
        Finally, to ensure no major bots are missed, all users with over 400 closed pull requests (slightly less than 500 users) have been manually checked, excluding four additional bots from the dataset.

\subsection{Contribution and Collaboration Metrics}
\label{sec:variable-construction}

    \revOneAdd{The following sections elaborate on the contribution and collaboration variables used to answer \rqn{1--4}.
    Table~\ref{tab:ecosystem-variables} provides an overview of these.}

    \subsubsection{Developer Contributions}
    
        Two answer \rqn{1,2,4}, developer contributions are calculated.
        Various studies have previously taken the number of pull requests submitted \citep{zhang_pull_2022, gousios_exploratory_2014, rastogi_relationship_2018, khadke_predicting_2012, zampetti_study_2019} or the number of code commits \citep{gousios_exploratory_2014, yu_determinants_2016, soares_acceptance_2015, soares_rejection_2015, khadke_predicting_2012, kononenko_studying_2018, zampetti_study_2019, zhang_pull_2022}, not considering the contributions done in alternative sources, like issue tracking systems or the comment section pull requests.
        % To limit the scope of this study, we do not include code commits.
        % However, issues have been included because non-coding contributions are prominent in open-source communities \citep{trinkenreich_hidden_2020, canovas_izquierdo_analysis_2021} but have not been formally studied yet.
        Table~\ref{tab:experience-types} overviews the contribution types included in this study.
        \revOneAdd{These metrics are calculated by counting the number of times contributors performed an activity.
        For example, the number of times they submitted a pull request or an issue.
        The pull request acceptance rate is counted by dividing the number of accepted pull requests by the total number of submitted pull requests.
        It is $0$ if the contributor submitted no pull requests.}
        
        Given a software ecosystem, developer contributions can be measured at two levels: intra-project and ecosystem-wide.
        Distinguishing between these two is important, as intra-project contributions have shown to be an important factor of pull request acceptance \citep{zhang_pull_2022, gousios_exploratory_2014, rastogi_relationship_2018, khadke_predicting_2012, zampetti_study_2019} and would confound the results if they were included in ecosystem contribution metrics.
        To prevent confounding, we define ecosystem-wide contributions as \qt{all contributions made in the ecosystem, excluding intra-project contributions.}
        More formally, given a focal project $p$ and a user $u$, ecosystem-wide contributions are the sum of all contributions made by $u$ in ecosystem projects $q$ minus the contributions made by $u$ in $p$.
        \revOneAdd{For example, if a user who previously submitted 1 pull request to project $A$, 3 to project $B$, and 5 to project $C$ (i.e., 9 in total) submits a new one to project $A$, the number of ecosystem pull requests is $9 - 1 = 8$.
        Conversely, the number of intra-project pull requests is 1.}

{

\begin{table}[!t]
    \centering
    \caption{Overview of contribution and collaboration types and the research questions they relate to}

    % Developer experience types.

    \begin{subtable}{\columnwidth}
        \subcaption{Contribution types (\rqn{1,2,4})}
        \begin{tabularx}{\columnwidth}{|l|X|}
            \hline
            \rowcolor{Gray} 
            \textbf{Name} & \textbf{Description} \\
            \hhline
            
            \textit{PR merge ratio} & 
                The fraction of accepted pull requests. \\
    
            \rowcolor{LighterGray}
            \textit{PRs submitted} & 
                The total number of submitted pull requests. \\
    
            \textit{PR comments} &
                The number of comments to a pull request. \\
    
            \rowcolor{LighterGray}
            \textit{Issues submitted} &
                The number of submitted issues. \\
    
            \textit{Issue comments} &
                The number of comments to an issue. \\
            
            \hline
        \end{tabularx}
        \label{tab:experience-types}
    \end{subtable}

    ~

    % Collaboration variables

    \begin{subtable}{\columnwidth}
        \subcaption{Collaboration types (\rqn{3,4})}
        \begin{tabularx}{\columnwidth}{|l|X|}
            \hline
            \rowcolor{Gray}
            \textbf{Name} & \textbf{Description} \\
            \hhline
            
            \textit{PR review} & One collaborator reviewed the pull request submitted by the other. \\
    
            \rowcolor{LighterGray}
            \textit{PR comment} & One collaborator commented on the pull request submitted by the other. \\
    
            \textit{PR discussion} & The collaborators participated in the pull request discussion. \\
    
            \rowcolor{LighterGray}
            \textit{Issue comment} & One collaborator commented on the issue submitted by the other. \\
    
            \textit{Issue discussion} & The collaborators participated in the issue discussion. \\
            
            \hline
        \end{tabularx}
        \label{tab:development-activity-types}
    \end{subtable}
\end{table}

}

    \subsubsection{Dependency Contributions}
    \label{sec:dependency-experience}
    
        Modern software commonly depends on other projects by reusing their functionalities \citep{decan_empirical_2019}.
        Dependencies are defined in two directions: downstream and upstream.
        A downstream dependency refers to a dependency from another project (a downstream project) to the focal project and an upstream dependency to the inverse.
        Consequently, ecosystem-wide contributions can be divided into upstream, downstream, and non-dependency contributions.
        Non-dependency contributions represent the contributions made to projects that are neither upstream nor downstream and make it possible to test the added benefit of dependencies.
        Calculating these contributions is equal to calculating ecosystem contributions, with the only difference being that they account for the (non-)existence of dependencies.
        
        Concretely, given a focal project $p$, downstream contributions are the sum of all contributions in ecosystem projects $q$ such that there exists a dependency from $q$ on $p$, upstream contributions such that there exists a dependency from $p$ on $q$, and non-dependency contributions such that neither exists.
        \revOneAdd{For example, given three projects $A$, $B$, and $C$ to which a user has submitted 1, 3, and 5 pull requests, respectively (again, 9 in total).
        If project $B$ depends on project $A$ and the developer submits a new pull request to project $A$, their downstream pull request count is 3, and their upstream count is 0 (as project $A$ does not depend on a project they contributed to).
        If the developer submits a new pull request to project $B$, their upstream pull request count is 1 (from project $A$), and their downstream count is 0 (as no projects they contributed to depend on project $B$).
        In both these cases, the non-dependency pull request count is 5 as that is the number of pull requests the user submitted to project $C$, which is not involved in a dependency with either $A$ or $B$.}
        
        The benefit of cascading dependencies \citep{geiger_labor_2021} has been reported in the past \citep{valiev_ecosystem-level_2018}.
        These are transitive dependencies from project $r$ on $p$ such that $r$ depends on an intermediary project $q$, which depends on $p$.
        In other words: $r$ depends on $q$ depends on $p$, thus $r$ transitively depends on $p$.
        However, we do not include it in our analysis because it is not obvious how it differs from regular dependencies due to limited evidence in the literature and the extra complexity it adds to our method.
        \revOneAdd{This could be addressed in future work.}
    
    \newcommand{\featureIntraproject}{Intra-project contributions}
\newcommand{\featureEcosystem}{Ecosystem contributions}
\newcommand{\featurenonDependency}{Non-dependency contributions}
\newcommand{\featureDownstream}{Downstream contributions}
\newcommand{\featureUpstream}{Upstream contributions}
\newcommand{\featureCommunity}{Ecosystem-wide community standing}
\newcommand{\featureCollaboration}{Direct collaboration}

{

\newlength{\Kwidth}
\setlength{\Kwidth}{0.25\hsize}
\newcolumntype{K}{>{\hsize=\Kwidth}X}
\newcolumntype{V}{>{\hsize=\dimexpr1\hsize-\Kwidth\relax}X}

% \newcolumntype{K}{>{\hsize=0.25\hsize}X}
% \newcolumntype{V}{>{\hsize=0.75\hsize}X}

\begin{table}[!t]
    \centering
    \small
    \caption{\revOneAdd{Overview of the contribution and collaboration variables used to answer \rqn{1--4}}}
    \revOneAdd{
    \begin{tabularx}{\columnwidth}{|K|V|}
        \hline
        \rowcolor{Gray}
        \textbf{Name} & \textbf{Description} \\
        \hhline
        \multicolumn{2}{|c|}{\textit{Contribution variables} (\rqn{1,2,4})} \\
        \hline
        \textit{\featureIntraproject} & 
            Past contributions to the same project that a new PR is submitted to. \\

        \rowcolor{LighterGray}
        \textit{\featureEcosystem} & 
            Past contributions to different projects (i.e., ecosystem projects) than the project to which a new PR is submitted.
            This is equal to the sum of the following ecosystem experience types. \\
        
        \textit{\featureDownstream} & 
            Past contributions to projects that depend on the project to which a new PR is submitted. \\

        \rowcolor{LighterGray}
        \textit{\featureUpstream} & 
            Past contributions to projects that the project to which a new PR is submitted is dependent. \\

        \textit{\featurenonDependency} & 
            Past contributions to ecosystem projects that the project to which a new PR is submitted is neither upstream nor downstream dependent. \\
            
        \hhline
        \multicolumn{2}{|c|}{\textit{Collaboration variables} (\rqn{3,4})} \\
        \hline
        \rowcolor{LighterGray}
        \textit{\featureCommunity} &
            The general connectedness in the ecosystem of the developer who submitted a new PR. \\
        
        \textit{\featureCollaboration} & 
            The number of times the submitter and integrator of a new PR have collaborated in the ecosystem. \\

        \hline
    \end{tabularx}
    }
    \label{tab:ecosystem-variables}
\end{table}

}

    \subsubsection{Ecosystem-wide Collaboration and Community Standing}
    \label{sec:social-network-analysis}
    
        Some studies already transcended project boundaries by studying the social connections between developers, suggesting that both general connectedness \citep{soto_analyzing_2017, yu_determinants_2016, zhang_pull_2022} and direct connectedness with your pull request integrator \citep{tsay_influence_2014, iyer_effects_2021, zhang_pull_2022} can affect pull request acceptance.
        These studies measure social connectedness using GitHub's follower feature.
        Although this provides insights into the social component of software engineering, it can hardly represent professional connectedness because following someone is accessible to anyone for any reason, not just professional reasons.
        Therefore, this study includes a new measure of connectedness: through collaboration in the ecosystem.
    
        Collaboration refers to the extent to which a contributor has worked with others, gaining experience and establishing trust with other developers.
        We consider connectedness in two manners: \textit{direct collaboration} and \textit{ecosystem-wide community standing}.
        Direct collaboration refers to the number of times two developers have directly cooperated.
        For example, by discussing something in an issue.
        Similarly, ecosystem-wide community standing is measured to account for the experience of the people they have collaborated with, as cooperating with many or well-experienced developers might be more valuable than working by yourself \revOneAdd{because having a higher number of connections might allow users to gain deeper knowledge of the system \citep{wang_survival_2012}.}
        We considered the five types of collaboration shown in Table~\ref{tab:development-activity-types}.
    
        We calculated direct collaboration and community standing using two metrics taken from social network analysis: \textit{link strength} and \textit{node centrality} \citep{schreiber_social_2020, newman_measures_2018}.
        Link strength refers to the number of collaborations connecting two contributors; i.e. direct collaboration.
        In turn, node centrality describes the general importance of a node in the network, which we use to measure community standing.
        Various well-known metrics for node centrality exist, like HITS, Eigenvector centrality, or PageRank \citep{newman_measures_2018}.
        Although these metrics are well-applicable to smaller datasets, their high computation time makes them impractical when applied with this study's intended level of granularity. %: applied to a 90-day window preceding each pull request; i.e., a sliding window.
        Therefore, we use an alternative: \textit{second-order degree centrality.}
        \revOneAdd{Appendix~\ref{app:so-degree} gives a formal definition of link strength and second-order degree centrality.}
    
        The intuition behind second-order degree centrality is that a contributor's centrality depends on the activity of the people they have collaborated with in the ecosystem.
        This metric grows when the collaborator either collaborated with many people who made few contributions, collaborated with few people who made many contributions, or somewhere in-between.
        This metric is inspired by previous works studying influential people in social networks \citep{chen_identifying_2012, brodka_analysis_2012} and trust in social networks \citep{stickgold_trust_2013}.
        A variant was also considered by \citet{jergensen_onion_2011} to identify core developers in OSS projects.%, however, it was excluded due to multicollinearity with other network variables.
        We exclude direct collaborations between developers as this has already been measured using link strength.
        In addition, we exclude any contributions of the neighbors that happened after the focal developer and their neighbor collaborated, as not accounting for time chronology would inflate the metric \citep{holme_map_2019}.
        Finally, similarly to ecosystem-wide contributions, we exclude neighbors' contributions to the focal project to prevent confounding by intra-project contributions.
        
        \revOneAdd{Because we extracted the five collaboration types shown in Table~\ref{tab:development-activity-types},} link strength and node centrality can be calculated for each.
        To reduce data analysis complexity, instead of calculating link strength and second-order degree centrality separately for each collaboration type, we aggregated the results using a weighted sum where the weights are inverse-proportional to the total number of collaborations \citep{kivela_multilayer_2014}.
        For example, because participating in the discussion in an issue-tracking system is more common than integrating someone's pull request, its weight is lower.
        % \revOneAdd{For a formal definition of link strength and second-order degree centrality, we refer to Appendix~\ref{app:so-degree}.}
        During preliminary analysis, experimentation was done using the analytical hierarchy process \citep{casola_ahp-based_2009} to fine-tune these weights.
        This is a systematic approach to manually assign weights to edges based on, for example, domain knowledge.
        Among others, we experimented based on the anticipated amount of information exchanged as, for example, the argument can be made that more information is exchanged in a comment on a pull request (giving feedback or providing context) than integrating it (clicking the accept/reject button).
        However, this step was omitted in the final methodology because no notable effect was observed in the results.

\subsection{Quantitative Analysis}
\label{sec:data-analysis}

    We used a combination of mixed-effects logistic regression and a random forest classifier to study the impact and predictive strength of the measured variables on pull request acceptance, respectively.
    Mixed-effects logistic regression describes the linear relationship between variables by calculating coefficients while respecting random effects across projects.
    This is important as some projects have higher pull request acceptance rates than others \citep{palyart_study_2018}.
    The control variables listed in Table~\ref{tab:control-variables} have been included to place the new variables in context.
    These variables are suggested for pull request research by \citet{zhang_pull_2022}, listing them as the principal factors for pull request decisions.
    We control for these variables by modeling them alongside the tested variables.
    Although \citeauthor{zhang_pull_2022} suggests including \qt{whether continuous integration (CI) is used,} we did not include it as the diversity of CI platforms makes it difficult to collect this data.
    We do not expect this to affect our results as \citeauthor{zhang_pull_2022} did not report that it correlates with any variable similar to the ones introduced in this work.
    Similarly, they suggest including \qt{whether the pull request submitter is a core member,} which is excluded because it is multicollinear with \qt{intra-project pull request merge ratio.}
    Further, although \qt{integrator experience} is included in the random forest models, it is excluded from the regression models as it is multicollinear with intra-project contributions.
    This is not a problem in random forest models.
    Finally, although \citeauthor{zhang_pull_2022} do not suggest \qt{whether the pull request submitter is a project newcomer,} we include it because it motivates asking \rqn{4}.
    This variable is calculated by tracing whether a pull request submitted by the user was accepted in the project before.
    
    To accurately answer the different research questions, we created multiple logistic regression models, simplifying the interpretation process: an \textit{ecosystem model} (\rqn{1}), a \textit{dependency model} (\rqn{2}), and a \textit{collaboration model} (\rqn{3}).
    In addition, to answer \rqn{4}, we created two additional models for each, distinguishing between the pull requests submitted by project newcomers and non-newcomers.

    The preliminary analysis showed that the top $2\%$ of the projects were responsible for $49\%$ of the pull requests, skewing the results significantly in favor of these projects.
    Therefore, a cap of 688 pull requests is put on the pull requests of these projects, matching the number of pull requests of the largest project that is not part of the top $2\%$.
    The pull requests in the top $2\%$ projects were sampled randomly.
    %We refer to \citet{lohr_sampling_2021} to underpin the representativeness of our samples per project, who describe a formula for sample size estimation for representative random samples.
    \revOneAdd{Using the sampling strategy described by \citet{lohr_sampling_2021}, we can calculate the minimum required sample size per project using a $5\%$ margin of error and a $95\%$ confidence interval.
    We cannot go into detail for each project as we sampled from 402 different projects.
    However, the project with the largest required sample only needs 95 pull requests for a representative sample.
    Increasing this number will only improve the sample’s representativeness.
    Because we sampled 688 pull requests for each project, our samples per project are representative.}
    
    \revOneAdd{Beyond project size, two other factors could skew our results: 1) extremely active contributors, who participate in many projects, and 2) users and organizations who own many projects.
    However, exploratory tests suggested they minimally impact our study's outcomes, minimizing this risk.
    Therefore, we do not actively account for these in our methodology.}
    
    Mixed effects logistic regression makes three assumptions: linearity between the variables and the log-odds ratio, absence of multicollinearity, and no strong outliers, as each harms the validity of the model results.
    Because most numerical values have a long-tail distribution, affecting the models' log-linearity, we used add-one log transformation $x' = \ln(x + 1)$.
    % To improve the log-linearity of the models, most numerical variables were transformed using add-one log transformation $x' = \ln(x + 1)$ because they follow a long-tail distribution.
    We used the add-one log transform instead of the normal log transform because all transformed features contain zeroes (for which the logarithm cannot be calculated).
    The transformed features are marked in Table~\ref{tab:coefficients-and-performance scores}.
    This tempers the impact of developers with a very high number of contributions.
    \revOneAdd{Multicollinearity is resolved using a \textit{combination} of the variance inflation factor ($VIF$) and the Spearman correlation coefficient $\rho$.
    For highly multicollinear variables ($VIF \geq 5$), we identified the variables that they correlated moderately to highly with ($\rho \geq 0.5$).
    % This highlighted various clusters of variables from which variables were manually picked and included in the statistical models.
    During analysis, it became apparent that the features in Table~\ref{tab:experience-types}, which are calculated for each of the contribution variables in Table~\ref{tab:ecosystem-variables}, correlate within their group.
    For example, the number of submitted intra-project pull requests correlates with the number of submitted intra-project issues, comments on issues, etc.
    This was also true for ecosystem contributions, downstream contributions, etc.
    Therefore, the logistic regression models only include one variable for each group.
    We selected the number of submitted issues for intra-project contributions and the number of submitted pull requests for all ecosystem contributions.
    We cannot use pull requests for intra-project contributions because, by definition, newcomers have submitted zero pull requests, whereas they might have submitted an issue.
    } % Variables that strongly correlated with another ($|\rho| \geq 0.5$) and are highly multicollinear ($VIF \geq 5$) were removed.
    
    Outliers were removed using Cook's distance \citep{cook_detection_2000}, filtering data points that affect regression coefficients disproportionally, using a cut-off threshold of  $4 / (n - k - 1)$, where $n$ is the number of observations, and $k$ is the number of predictors.
    This removed between $0.1\%$ and $2.6\%$ data points across models.
    \revOneAdd{Finally, all numerical data was normalized using min-max normalization, simplifying the comparison between variables as it transforms their range to the same scale.
    Although this method is sensitive to outliers, our results are minimally affected because we applied the log transformation and removed outliers.}
    
    \revOneAdd{Table~\ref{tab:control-variables} provides an overview of the included control variables.
    Table~\ref{tab:ecosystem-variables} provides an overview of the included contribution variables used to answer \rqn{1--4}.}
    \revOneAdd{We calculated each variable using a snapshot of 90 days before the pull request was closed because modern software changes rapidly over time.
    This might make experience acquired in a project a longer time ago less relevant.
    This specific window is a trade-off between the amount of considered data and the computational complexity of social network analysis metrics.
    Through experimentation, we found that smaller time windows created problematically sparse datasets while previous work \citep{meneely_socio-technical_2011} has suggested expanding beyond 90 days yields limited additional insights.
    It should be noted that the feature \qt{PR submitter is a newcomer} is exempt from this 90-day window.
    If developers successfully submit a pull request at any point, they will never be considered newcomers to that project again.
    } % To account for the fact that software changes rapidly over time, each variable is calculated using a snapshot of 90 days before the pull request is closed.

    Finally, to identify the predictive strength of our features, random forest is used \citep{breiman_random_2001, pedregosa_scikit-learn_2011}.
    To measure the importance of individual features, the mean decrease in Gini is calculated, which estimates the amount of information lost when a predictor is removed \citep{breiman_random_2001, pedregosa_scikit-learn_2011}.
    \revOneAdd{Random forest is appropriate for our study because it allows us to evaluate features' predictive strength in detail using a simple yet commonly applied ML algorithm.
    For example, we can include all our metrics instead of a subset because we do not have to account for multicollinearity explicitly.
    In addition, random forest is more robust against overfitting \citet{breiman_random_2001}, improving the representativeness of its feature importance estimates as they are less sensitive to individual data points.
    Random forests can be affected by random effects, which can potentially inflate feature importance. However, we find very limited evidence of this in our context because the results reported by random forest align with those reported by mixed-effects logistic regression (we discuss this in Section~\ref{sec:results}).}
    
    To evaluate the predictive strength of groups of variables, an \textit{inverse ablation} study was performed (i.e., separate models were trained using subsets of features) and evaluated using F1 scores.
    \revOneAdd{We consider seven variable groups: one for the control variables in Table~\ref{tab:control-variables}, five separate groups for the contribution variables in Table~\ref{tab:ecosystem-variables}, and one for the collaboration variables in Table~\ref{tab:ecosystem-variables}.
    Each group with contribution variables contains the five features described in Table~\ref{tab:experience-types}.
    }

\definecolor{Gray}{gray}{.9}

\begin{table}[!t]
    \centering
    \caption{\revOneAdd{Example of the qualitative study's summarization and classification process using a reference from a pull request to an issue in another project, including the focal pull request's title and description, our summary, and the encoding of that summary}}
    \revOneAdd{
    \begin{tabularx}{\textwidth}{|l|X|}
        \hline
        \textbf{Focal pull request} & \href{https://github.com/TypeStrong/ts-loader/pull/774}{TypeStrong/ts-loader\#774} \\
        \textbf{Referenced activity} & \href{https://github.com/microsoft/TypeScript/issues/12358}{microsoft/TypeScript\#12358} \\
        \hline\hline
        \textbf{Pull request title}         & Fix dependency resolution when using pnpm \\
        \hline
        \textbf{Pull request description}               
                                            & $\bullet$~Add realpath to moduleResolutionHost so that TypeScript could follow symlink as it supposed to be \\
                                            & $\bullet$~Now that typescript follow symlink, npmLink test fails in 2.8.1 due to \textit{``Guidance on shipping ts within node\_modules''} (\href{https://github.com/microsoft/TypeScript/issues/12358}{\underline{microsoft/TypeScript\#12358}}), fixed by modifying tsconfig file. (Also related to Add \textit{``allowTsInNodeModules option for importing .ts files from node\_modules.''} (\#773)) \\
                                            & $\bullet$~Created a new test to cover this scenario. \\
                                            & \\
                                            & Fix \#768 \\
        \hline\hline
        \textbf{Summary} & References issue to describe a failing test and as motivation to make the corresponding changes to prevent that from happening. \\
        \hline
        \textbf{Codes}  & $\bullet$~Motivation: feature/problem/bug description \\
                        & $\bullet$~Outgoing reference \\
        \hline
         
    \end{tabularx}
    \label{tab:classification-example}
    }
\end{table}

\revOneAdd{

\subsection{Exploratory Qualitative Analysis}\label{sec:quaitative-analysis}

    To further corroborate our findings, we complement the results of our quantitative study with an exploratory qualitative analysis.
    An exploratory study fits our situation as, to our knowledge, no work has previously addressed ecosystem participation qualitatively by analyzing pull requests.
    Therefore, an exploratory study like this one lays the necessary groundwork for future work to address this in detail.
    % Our results suggest various of these.
    
    We performed a \textit{two-stage analysis} to identify why ecosystem participation positively impacts pull request decisions using inductive thematic analysis \citep{mayring_qualitative_2014}, following the grounded theory framework \citep{seaman_qualitative_1999}.
    The themes are then used to encode the studied pull requests, enabling us to analyze the outcomes statistically.
    To do this, we first summarized a subset of pull requests to capture \textit{general behaviors} while abstracting technical details.
    From these, we extracted codes and refined them iteratively until applying them became (almost) trivial with minimal loss of nuance.
    Once this set of codes was well-defined, we classified the remaining data points without summarizing them.
    Table~\ref{tab:classification-example} gives an example of this process.

    Because experience has previously been linked with contribution types \citep{jergensen_onion_2011}, we first analyze a sample of pull requests on the types of contributions made in them.
    We differentiate between \textit{``content contributions''} and \textit{``meta contributions.''}
    Content contributions are defined as any pull request that introduces new features, adds test cases, fixes bugs/problems, changes documentation, or changes the configuration of a project (i.e., the content).
    Conversely, meta contributions are defined as pull requests that make changes like repository configurations (e.g., the gitignore), CI pipelines, project releases, or dependency updates as part of maintenance tasks.
    Content contributions are different from meta contributions in the sense that meta contributions do not change the functionality or description of the product itself.
    Arguably, dependencies should be considered separately.
    Therefore, we differentiate between dependency changes as part of e.g., introducing a new feature (a content contribution) and version bumps as part of a maintenance task (a meta contribution).

    Because our study’s primary interest is ecosystem collaboration, we marked pull requests with clear ecosystem involvement.
    This is generally indicated by developers talking indirectly about other projects, by a project being referenced directly, or by another development activity referencing the focal pull request.
    GitHub provides a feature for the last one, making them easy to find.
    To get a better understanding of ecosystem involvement in pull requests, we analyze the motivations of ecosystem references.
    The expectation here is that developers’ reasons for referencing other projects might change when their involvement in the ecosystem changes.
    For example, developers from downstream projects might highlight issues posted in those projects more frequently.

    Because no similar previous study has been performed in this domain, we took a stratified sample from our pull request dataset.
    For each of the ecosystem experience metrics (ecosystem, upstream, downstream, community standing, and direct collaboration), we sampled 50 pull requests.
    We then sampled another 50 pull requests from all strata to emphasize pull requests submitted by newcomers.
    We did not consider non-dependency experience separately as it overlaps highly with the ecosystem stratum (this is by design).
    Finally, we sampled an additional 50 pull requests submitted by people with intra-project contributions.
    To study the differences between submitters who score higher on these metrics versus those who score lower, instead of sampling randomly, we divided the pull requests of each metric into four quartiles and upper outliers (i.e., pull requests submitted by people who score very high compared to non-outliers) and sampled 10 pull requests from each of these or fewer if not enough pull requests were available in that group (this was only the case for people with an extreme amount of upstream experience).
    Further, we removed pull requests if they were no longer accessible on GitHub (4 cases).
    In total, we sampled 538 pull requests spread across 11 strata.
    Because this sampling strategy is quite specific, the reported results are not representative of the general NPM ecosystem.
    However, they give an early description of this study’s quantitative outcomes, laying the groundwork for future work.

    Using these classifications --- content versus meta contributions and motivations for referencing another project --- we performed a series of statistical analyses to identify whether there is a relationship between pull requests submitted by developers with different amounts and types of ecosystem experience.
    We use tests of independence, which test the relationship between categorical variables.
    We use two statistical tests because we have fairly high-dimensional data (due to the number of strata and labels).
    We use Chi-squared, $\chi^2$, when comparing multinomial data where each of the observed classes has 5 or more observations (a prerequisite of the model to perform well), and use Fisher’s Exact Test, $F$, when this is not the case.
    The latter, however, restricts us to $2\times2$ comparisons only.
    Because the number of observations in various of the Fisher’s Exact Tests were zero, which prevented us from calculating the odds ratio, we applied additive smoothing, $F^*$, meaning 1 was added to all observations in the test.
    Although this generally makes the method more conservative, it did not affect our results or conclusions meaningfully.
    This ratio is important as it tells us how strong the relationship is between the tested variables.
    For clarity, we report the test used ($\chi^2$, $F$, $F^*$) in tandem with the significance value~($p$) and the odds ratio.
    
    To give an example, our statistical tests showed that people with lower intra-project contributions (those in the bottom two quartiles) tend to make content contributions more commonly ($F^*$, $p=0.046$, $odds=8.4$).
    The odds ratio of $8.4$ suggests that, in our data, they were $8.4$ times more likely to make a content contribution versus not making one.
    Complementary, we also found weak evidence that they make fewer meta contributions ($F^*$, $p=0.095$, $odds=0.127$).
    The odds ratio of $0.127$ means they were $0.127$ times more likely to make meta contributions; i.e., people with more intra-project contributions (those in the top two quartiles) are $\frac{1}{0.127}=7.874$ more likely to make meta contributions.
    This suggests that meta contributions are reserved for more mature contributors.
    
}

\section{Results}
\label{sec:results}

\newcommand{\cmark}[1]{\hphantom{$^\star$}#1$^\star$}

% Commands for formatting cell entries
\def\stripzero#1{\expandafter\stripzerohelp#1}
\def\stripzerohelp#1{\ifx 0#1\expandafter\stripzerohelp\else#1\fi}

\newcommand{\PadIfPos}[1]{\ifthenelse{\NOT\equal{#1}{-}}{\hphantom{$-$}}{}}

\def\IsSignificant#1{TT\fi\ifcat_\ifdim#1pt<0.006pt _\else A\fi}
\newcommand{\Sig}[1]{\if\IsSignificant{#1}^{*}\fi}

\newcommand{\Res}[4]{\PadIfPos{#1}$#1\stripzero{#2}\Sig{#3}(\stripzero{#4})$}

\newcommand{\NoRes}{\textit{n/a}}

\definecolor{Gray}{gray}{.9}
\definecolor{LightGray}{gray}{.92}

\afterpage{
\begin{landscape}
\begin{table}

\newcolumntype{K}{>{\hsize=0.2\hsize}X}
\newcolumntype{V}{>{\hsize=0.26\hsize}Y}
\newcolumntype{W}{>{\hsize=0.12\hsize}Y}

\centering
\small

\caption{Overview of the coefficients (\textit{coef.}), standard error (\textit{std. err.}) calculated by the mixed-effects logistic regression models, containing control, contribution (contr.), and collaboration variables. The sign of the coefficient indicates the relationship type (negative vs. positive). $^\diamond$Add-one log-transformed variable; $^* p < .001$}

\begin{subtable}{\linewidth}
    \caption{Models that used all data points, used to answer \rqn{1--3}}
    \label{tab:coefficients-and-performance scores-general}
    
    \adjustbox{width=\linewidth}{

        \begin{tabularx}{\linewidth}{|K|V|V|V|}
            \cline{2-4}
        
            \multicolumn{1}{c|}{} & 
                \textbf{Ecosystem model (\rqn{1})} &
                \textbf{Dependency model (\rqn{2})} &
                \textbf{Collaborative model (\rqn{3})} \\
            
            \chline{2-4}{4}
            
            \rowcolor{Gray} 
            \textbf{Control variables} &
                \textit{Coef. (std. err.)} & 
                \textit{Coef. (std. err.)} & 
                \textit{Coef. (std. err.)}  \\
            \hline
            
            \textit{\featureNewcomer} & 
                \Res{-}{0.129}{0.000}{0.001} &
                \Res{-}{0.129}{0.000}{0.001} &
                \NoRes \\
                
            \rowcolor{LightGray}
            \textit{\featureSelfIntegrated} &
                \Res{-}{0.262}{0.000}{0.001} &
                \Res{-}{0.261}{0.000}{0.001} &
                \NoRes \\
            
            \textit{\featureComments} &
                \Res{-}{0.127}{0.000}{0.001} &
                \Res{-}{0.127}{0.000}{0.001} &
                \Res{-}{0.118}{0.000}{0.001} \\
            
            \rowcolor{LightGray}
            \textit{\featureReference} &
                \Res{}{0.037}{0.000}{0.001} &
                \Res{}{0.037}{0.000}{0.001} &
                \Res{}{0.040}{0.000}{0.001} \\
            
            \textit{\featureExternalComments} &
                \Res{}{0.030}{0.000}{0.001} &
                \Res{}{0.030}{0.000}{0.001} &
                \Res{-}{0.006}{0.000}{0.001} \\
            
            \rowcolor{LightGray}
            \textit{\featureAge}$^\diamond$ & 
                \Res{-}{0.482}{0.000}{0.002} &
                \Res{-}{0.478}{0.000}{0.002} &
                \Res{-}{0.435}{0.000}{0.002} \\
            
            \textit{\featureCommits}$^\diamond$ &
                \Res{}{0.189}{0.000}{0.005} &
                \Res{}{0.186}{0.000}{0.005} &
                \Res{}{0.124}{0.000}{0.005} \\
            
            \hhline
            
            \rowcolor{Gray} 
            \textbf{Contribution variables} & 
                \textit{Coef. (std. err.)} & 
                \textit{Coef. (std. err.)} & 
                \textit{Coef. (std. err.)}  \\
            \hline
            
            \textit{Intra-project contr.}$^\diamond$ & 
                \Res{}{0.272}{0.000}{0.003} &
                \Res{}{0.274}{0.000}{0.003} &
                \Res{}{0.202}{0.000}{0.003} \\
            
            \rowcolor{LightGray}
            \textit{Ecosystem contr.}$^\diamond$ &
                \Res{}{0.252}{0.000}{0.002} &
                \NoRes &
                \NoRes \\
    
            \textit{Non-dependency contr.}$^\diamond$ & 
                \NoRes &
                \Res{}{0.215}{0.000}{0.002} &
                \NoRes \\
    
            \rowcolor{LightGray}
            \textit{Downstream contr.}$^\diamond$ & 
                \NoRes &
                \Res{}{0.128}{0.000}{0.005} &
                \NoRes \\
            
            \textit{Upstream contr.}$^\diamond$ & 
                \NoRes &
                \Res{}{0.097}{0.000}{0.006} &
                \NoRes \\
            
            \rowcolor{LightGray}
            \textit{Eco. community standing}$^\diamond$ &
                \NoRes &
                \NoRes &
                \Res{}{0.167}{0.000}{0.004} \\
            
            \textit{Direct collaboration$^\diamond$} &
                \NoRes &
                \NoRes &
                \Res{}{0.322}{0.000}{0.004} \\
              
            \hline
        \end{tabularx}
    }
\end{subtable}

~\\~\\
%%%%%%%%%%%%%%%%%%%%%%%%%
% BEGIN newcomer / non-newcomer data
%%%%%%%%%%%%%%%%%%%%%%%%%

\begin{subtable}{\linewidth}

    \caption{Models distinguishing between project newcomer and non-newcomer data points, used to answer \rqn{4}}
    \label{tab:coefficients-and-performance scores-ftc}
    \adjustbox{width=\linewidth}{

        \begin{tabularx}{\linewidth}{|K|W|W|W|W|W|W|}
        
            \cline{2-7}
            \multicolumn{1}{c|}{} &
                \multicolumn{2}{c|}{\textbf{Ecosystem model}} &
                \multicolumn{2}{c|}{\textbf{Dependency model}} &
                \multicolumn{2}{c|}{\textbf{Collaborative model}} \\
            \cline{2-7}
            \multicolumn{1}{c|}{} &
                \textit{Newcomer} & \textit{Non-newcomer} &
                \textit{Newcomer} & \textit{Non-newcomer} &
                \textit{Newcomer} & \textit{Non-newcomer} \\
        
            \chline{2-7}{7}
            
            \rowcolor{Gray} 
            \textbf{Control variables}
                & \textit{Coef. (std. err.)}
                & \textit{Coef. (std. err.)}
                & \textit{Coef. (std. err.)}
                & \textit{Coef. (std. err.)}
                & \textit{Coef. (std. err.)}
                & \textit{Coef. (std. err.)} \\
            \hline
        
            \textit{\featureSelfIntegrated} & 
                \Res{-}{0.562}{0.000}{0.002} &
                \Res{-}{0.125}{0.000}{0.001} &
                
                \Res{-}{0.563}{0.000}{0.002} &
                \Res{-}{0.125}{0.000}{0.001} &
                
                \NoRes &
                \NoRes \\
        
            \rowcolor{LightGray}
            \textit{\featureComments} & 
                \Res{-}{0.156}{0.000}{0.002} &
                \Res{-}{0.116}{0.000}{0.001} &
                
                \Res{-}{0.156}{0.000}{0.002} &
                \Res{-}{0.117}{0.000}{0.001} &
                
                \Res{-}{0.043}{0.000}{0.002} &
                \Res{-}{0.096}{0.000}{0.001} \\

            \textit{\featureReference} &
                \Res{}{0.044}{0.000}{0.001} &
                \Res{}{0.032}{0.000}{0.001} &
                
                \Res{}{0.044}{0.000}{0.001} &
                \Res{}{0.032}{0.000}{0.001} &
                
                \Res{}{0.048}{0.000}{0.002} &
                \Res{}{0.032}{0.000}{0.001} \\

            \rowcolor{LightGray}
            \textit{\featureExternalComments} &
                \Res{}{0.025}{0.000}{0.002} &
                \Res{}{0.047}{0.000}{0.001} &
                
                \Res{}{0.024}{0.000}{0.002} &
                \Res{}{0.047}{0.000}{0.001} &
                
                \Res{-}{0.066}{0.000}{0.002} &
                \Res{}{0.028}{0.000}{0.001} \\

            \textit{\featureAge}$^\diamond$ & 
                \Res{-}{0.503}{0.000}{0.003} &
                \Res{-}{0.494}{0.000}{0.002} &
                
                \Res{-}{0.500}{0.000}{0.003} &
                \Res{-}{0.490}{0.000}{0.002} &
                
                \Res{-}{0.202}{0.000}{0.003} &
                \Res{-}{0.458}{0.000}{0.002} \\

            \rowcolor{LightGray}
            \textit{\featureCommits}$^\diamond$ &
                \Res{}{0.216}{0.000}{0.010} &
                \Res{}{0.282}{0.000}{0.005} &
                
                \Res{}{0.213}{0.000}{0.010} &
                \Res{}{0.279}{0.000}{0.005} &
                
                \Res{-}{0.140}{0.000}{0.012} &
                \Res{}{0.204}{0.000}{0.005} \\
        
            \hhline
        
            \rowcolor{Gray} 
            \textbf{Contribution variables}
                & \textit{Coef. (std. err.)}
                & \textit{Coef. (std. err.)}
                & \textit{Coef. (std. err.)}
                & \textit{Coef. (std. err.)}
                & \textit{Coef. (std. err.)}
                & \textit{Coef. (std. err.)} \\
            \hline

            \textit{Intra-project contr.}$^\diamond$ & 
                \Res{}{0.213}{0.000}{0.013} &
                \Res{}{0.169}{0.000}{0.003} &
                
                \Res{}{0.214}{0.000}{0.013} &
                \Res{}{0.170}{0.000}{0.003} &
                
                \Res{}{0.186}{0.000}{0.015} &
                \Res{}{0.109}{0.000}{0.003} \\

            \rowcolor{LightGray}
            \textit{Ecosystem contr.}$^\diamond$ &
                \Res{}{0.376}{0.000}{0.004} &
                \Res{}{0.164}{0.000}{0.002} &
                
                \NoRes &
                \NoRes &
                
                \NoRes &
                \NoRes \\

            \textit{Non-dependency contr.}$^\diamond$ &
                \NoRes &
                \NoRes &
                
                \Res{}{0.324}{0.000}{0.005} &
                \Res{}{0.133}{0.000}{0.002} &
                
                \NoRes &
                \NoRes \\
    
            \rowcolor{LightGray}
            \textit{Downstream contr.}$^\diamond$ & 
                \NoRes &
                \NoRes &
                
                \Res{}{0.295}{0.000}{0.016} &
                \Res{}{0.094}{0.000}{0.004} &
                
                \NoRes &
                \NoRes \\

            \textit{Upstream contr.}$^\diamond$ & 
                \NoRes &
                \NoRes &
                
                \Res{}{0.318}{0.000}{0.013} &
                \Res{}{0.071}{0.000}{0.006} &
                
                \NoRes &
                \NoRes \\
        
            \rowcolor{LightGray}
            \textit{Eco. community standing}$^\diamond$ &
                \NoRes &
                \NoRes &
                
                \NoRes &
                \NoRes &
                
                \Res{}{0.335}{0.000}{0.008} &
                \Res{}{0.071}{0.000}{0.004} \\

            \textit{Direct collaboration}$^\diamond$ &
                \NoRes &
                \NoRes &
                
                \NoRes &
                \NoRes &
                
                \Res{}{0.517}{0.000}{0.012} &
                \Res{}{0.240}{0.000}{0.004} \\
        
            \hline
        \end{tabularx}
    }
\end{subtable}

\label{tab:coefficients-and-performance scores}

\end{table}

\end{landscape}
}

The collected dataset consists of 1.8 million pull requests and 2.1 million issues spread across 20,052 projects, submitted by 190,898 unique users, and spread across 9 years.
Of these, $72\%$ of the projects were collected using the popularity criterion and $28\%$ with the dependency criterion.
The latter was almost equally split into upstream and downstream projects.
Of these projects, 19,609 ($98\%$) have an upstream dependency and 8,688 ($43\%$) have a downstream dependency on another project in the dataset, which could suggest a core-periphery structure in the dependency network (e.g., \citet{barabasi_emergence_1999}).
\revOneAdd{Project newcomers submitted 535,398 pull requests ($29.48\%$), highlighting their prominence in open-source software ecosystems.}
As described in Section~\ref{sec:data-analysis}, this dataset was sampled according to the number of pull requests per project to improve internal validity, for which 1.2 million data points ($66\%$) were used for inference.

\subsection{Quantitative Analysis}
\label{sec:res-quantitative-data-analysis}

Table~\ref{tab:coefficients-and-performance scores} presents the results of the mixed-effects logistic regression models, showing the general case and differentiating between pull requests submitted by newcomers and non-newcomers.
% During analysis, it became apparent that all contribution types listed in Table~\ref{tab:experience-types} are positively correlated.
% This means that for ecosystem-wide, upstream, downstream, non-dependency, and intra-project contributions, all contribution variables shown in Table~\ref{tab:experience-types} had a Spearman coefficient greater than $0.5$.
% Therefore, the logit models only include one variable for each group.
The results show that each measured variable significantly and positively correlates with pull request acceptance.
For example, looking at \revOneAdd{\qt{ecosystem contributions}} in Table~\ref{tab:coefficients-and-performance scores-general}, a pull request submitted by someone who submitted many pull requests in the ecosystem is up to $25.2\%$ more likely to be accepted (as indicated by its coefficient $0.252$).

\begin{figure}[!b]
    \centering
    \includegraphics[width=0.7\columnwidth]{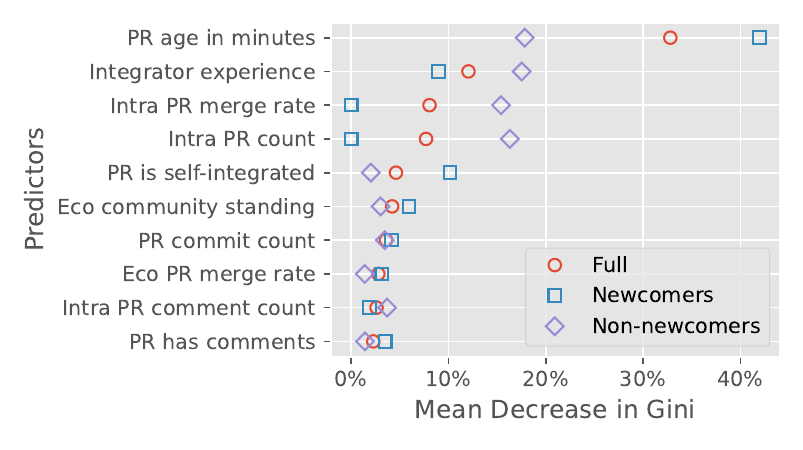}
    \caption{Feature importance plot showing the mean decrease in Gini calculated using random forest trained using the full dataset and the (non-)newcomer subsets}
    \label{fig:feature-importance}
\end{figure}

To identify the importance of the used features, Figure~\ref{fig:feature-importance} overviews the importance of the top 10 features of the random forest models.\footnote{Refer to the replication package for an overview of all 35 variables.}
It stands out that the majority of information ($56\%$) is acquired from the control variables: \textit{PR lifetime} ($32\%$) \textit{integrator experience} ($12\%$), \textit{intra-project PR count} ($8\%$) and \textit{PR is self-integrated} ($4\%$).
This is unsurprising, as previous work \citep{zhang_pull_2022} listed them as principal factors of pull request decisions.
Conversely, the absence of the other control variables stands out as it suggests some of our features yield more information, like \textit{node centrality} ($4\%$) and \textit{pull requests submitted in the ecosystem} ($3\%$).

\revOneAdd{Although random forests can be affected by random effects across projects, we find very limited evidence of this in our context.
This is because features with high impacts in the mixed-effects logistic regression model also have high feature importance.
The only exceptions here are \qt{PR submitter is a newcomer} and \qt{direct collaboration.}
The limited importance of direct collaboration is due to its sparsity as this is non-zero for only $10\%$ of the pull requests.
Testing for this subset of data significantly increased its importance, becoming the seventh most important feature, just below ecosystem-wide community standing.
The limited importance of being a newcomer is likely because pull requests submitted by newcomers are generally harder to predict, such that only $65.8\%$ of the pull requests submitted by newcomers are accepted, compared to $85.7\%$ for non-newcomers.}

To compare the importance of the different contribution levels (intra-project, ecosystem, etc.) in the random forest models, we performed a pair-wise comparison using the feature importance scores of their comprising contribution types (pull requests submitted, etc.).
Here, we observe whether the feature importance of \textit{all} contribution types of a contribution level is higher than those in another, suggesting that the contribution level is of higher importance.
We found that intra-project contributions are more important than ecosystem contributions, downstream contributions are more important than upstream contributions, and non-dependency contributions are more important than upstream and downstream contributions.
Interestingly, only the last of these three applied to project newcomers, suggesting that the added benefit downstream and intra-project contributions are relatively less evident when predicting the outcome of their pull requests.

\definecolor{Gray}{gray}{.9}
\definecolor{LightGray}{gray}{.92}

\begin{table}[!t]

\small
\centering

\caption{The mean and standard deviation of the F1 scores calculated with the random forest models, trained through inverse ablation using variable groups, calculated using 5-fold cross-validation. The models are trained using all data and the (non-)newcomer subsets. The baseline is the probability of observing a merged pull request}
\begin{tabularx}{\columnwidth}{|l|Y|Y|Y|}

    \hline
    \rowcolor{Gray}
    \textbf{Variable group} & \textbf{All data} & \textbf{Newcomer} & \textbf{Non-newcomer} \\
    \hhline

    \textit{Baseline}       & .787          & .658          & .857 \\
    \hhline

    \rowcolor{LightGray}
    \textit{Control}        & .877~(.0004) & .796~(.0009) & .921~(.0003) \\
    
    \textit{Intra-project contributions}  & .881~(.0003) & .793~(.0006) & .925~(.0007) \\
    \hline

    \rowcolor{LightGray}
    \textit{Combined}       & .910~(.0002) & .807~(.0009) & .957~(.0003) \\
    \hhline

    \textit{Ecosystem contributions}      & .877~(.0002) & .782~(.0011) & .920~(.0003) \\
    
    \rowcolor{LightGray}
    \textit{Non-dependency contributions} & .877~(.0001) & .783~(.0010) & .920~(.0005) \\
    
    \textit{Downstream contributions}     & .880~(.0005) & .793~(.0011) & .923~(.0006) \\
    
    \rowcolor{LightGray}
    \textit{Upstream contributions}       & .880~(.0007) & .793~(.0014) & .923~(.0005) \\
    
    \textit{Collaboration variables}  & .864~(.0006) & .759~(.0011) & .911~(.0002) \\
    \hline
    
    \rowcolor{LightGray}
    \textit{Combined}       & .874~(.0003) & .773~(.0013) & .920~(.0006) \\
    \hhline

    \textbf{\textit{All variables}}     & .921~(.0002) & .830~(.0005) & .959~(.0004) \\
    \hline

\end{tabularx}
\label{tab:model-performance}

\end{table}

Table~\ref{tab:model-performance} provides the broadest view of the analysis, showing the results of the reverse ablation study.
These results include a baseline for each model, which equals the probability of observing a merged pull request in the dataset, as pull requests are more commonly accepted than refused.
Although any model with an F1 score over $0.5$ (intuitively, 50\% correct classifications) would have some predictive power, only models outperforming this baseline score better than a probabilistic guesser.
Looking at Table~\ref{tab:model-performance}, we see that each group performs almost equally, reaching F1 scores of approximately $0.88$, outperforming the baseline, with almost no deviation across data folds used for cross-validation.
Models trained using only collaboration metrics perform slightly worse, scoring $1.3$-$1.6\%$ lower.

Interestingly, although the model including all intra-project variables improves these separate models between $1.3\%$ and $3.6\%$, creating a model that uses the newly included ecosystem variables did not improve classification performance compared to the models trained with their comprising sub-groups.
Regardless, combining all variables improves classification by $0.2\%$ to $2.3\%$ \revOneAdd{compared to the model only considering control variables and intra-project contributions}, especially for pull requests submitted by project newcomers.
This suggests that classifying pull requests submitted by this group benefits from ecosystem-wide information.

\revOneAdd{

\definecolor{Gray}{gray}{.9}
\definecolor{LightGray}{gray}{.92}

\begin{table}[!t]
    \centering
    \caption{\revOneAdd{Overview of the reasons to reference other ecosystem projects}}
    \revOneAdd{
    \begin{tabularx}{\textwidth}{|l|X|}
        \hline
        \rowcolor{Gray} 
        \textbf{Name (acronym)} & \textbf{Definition} \\
        \hline\hline
        
        \rowcolor{Gray} 
        \multicolumn{2}{|c|}{\textit{Motivation}} \\
        \hline
        Change accommodation (MC) & \textit{The reference introduces or describes a change that must be accounted for by the focal project (e.g., a changed API interface).} \\
        \rowcolor{LightGray}
        Problem description (MD) & \textit{The reference describes or illustrates a problem/bug, or a new feature/change proposal.} \\
        \hline\hline
        
        \rowcolor{Gray} 
        \multicolumn{2}{|c|}{\textit{Solution}} \\
        \hline
        Specification (SS) & \textit{The reference provides a natural language specification of the implemented solution.} \\
        \rowcolor{LightGray}
        Used in change (SU) & \textit{The reference contains the solution introduced/integrated in the focal pull request.} \\
        (Partial) implementation (SI) & \textit{The reference implements (part of) the solution, is a very similar one, or implements a strongly related change that tackles a similar problem as the focal pull request.} \\
        \rowcolor{LightGray}
        Affected artifacts (SA) & \textit{The reference is to an artifact (e.g., code), indicating that it will be changed as part of the proposed solution.} \\
        \hline\hline

        \rowcolor{Gray} 
        \multicolumn{2}{|c|}{\textit{Other}} \\
        \hline
        System description (OD) & \textit{The reference describes the general architecture of the project or a design decision. The reference provides context.} \\
        \rowcolor{LightGray}
        Change consequence (OC) & \textit{The reference indicates the effect of the proposed change on other projects or the focal project itself (because the pull request does address them).} \\
        Convention reference (OR) & \textit{The reference specifies a coding or pull request convention used in the project.} \\
        \rowcolor{LightGray}
        Other (OO) & \textit{References of which the motivation is unclear or cannot be captured by the other classes.} \\
        \hline
         
    \end{tabularx}
    \label{tab:ref-motivations}
    }
\end{table}

\subsection{Exploratory Qualitative Analysis}
\label{sec:res-qualitative-data-analysis}

    We complement our quantitative analysis by classifying pull requests based on the type of contribution they make, marking them as \qt{content contributions} when they change the content of a project (e.g., new features, test cases, or fixes to bugs/problems), and \qt{meta contributions} when they do not (e.g.,  repository configurations like the gitignore, or CI pipelines).
    In total, $368$ pull requests ($68.53\%$) make content contributions, $200$ pull requests ($37.24\%$) make meta contributions, and they do both in a limited number of cases ($15$; $4\%$).
    % In a limited number of cases ($15$), they do both.
    We found $111$ ($20.63\%$) pull requests with clear ecosystem involvement (i.e., they either reference another project or are referenced by another project).
    Zooming into these, we identified $202$ references to or from external projects, of which $156$ ($77.23\%$) were references from the focal pull request to something else, and $46$ ($22.23\%$) were references from another project to the focal pull request.
    Most pull requests are associated with one reference ($67$), some with two ($26$), and a few with three or more ($18$).
    Of all references, $175$ ($86.63\%$) were from/to another GitHub project.
    The other $27$ references ($13.37\%$) were indirect (not specifying who they were talking about), to project websites (e.g., blogs or documentation), or external issue trackers (e.g., Jira).
    Of the GitHub references other GitHub projects, $96$ ($54.9\%$) were to/from a project owned by the same organization (e.g., CKEditor, CanJS, or Microsoft).
    % $155$ references are outgoing (i.e., the focal pull request references something else), and $46$ are incoming (i.e., another development activity in GitHub references the focal pull request).

    We were able to identify three overarching reasons to reference another project: 1) motivation to make a change ($78$ cases), 2) solution description ($94$ cases), and 3) another reason ($34$ cases).
    These were subdivided into the 10 specific reasons described in Table~\ref{tab:ref-motivations}.
    For example, a bug reported in another project might have to be patched in a downstream project, thus motivating the change proposed in a new pull request.
    Similarly, in some cases, other projects have already resolved a problem the focal project has.
    Those projects are referenced as inspiration for the solution proposed in a new pull request.

\begin{figure}[!b]
    \centering
    \includegraphics[width=0.8\linewidth]{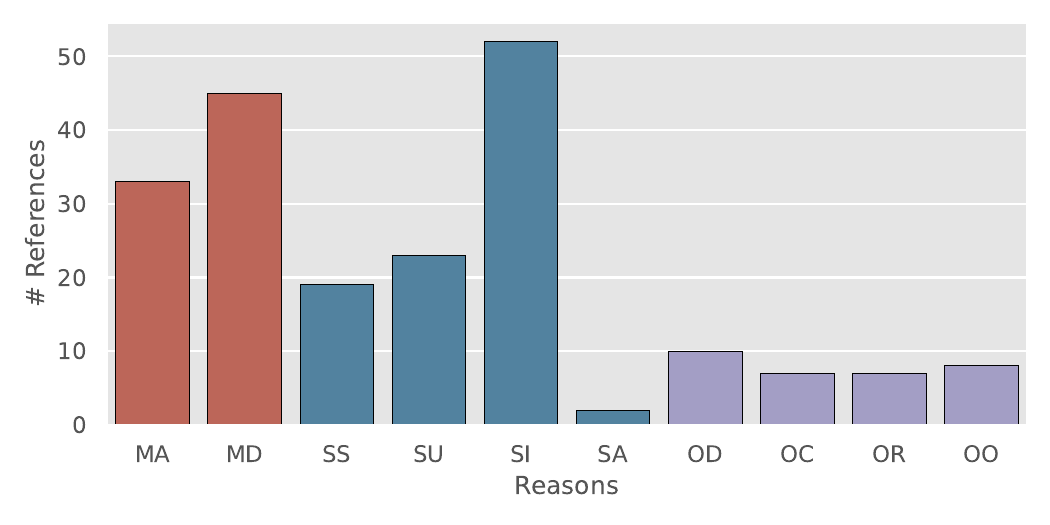}
    \caption{\revOneAdd{Overview of the reasons to reference another project}}
    \label{fig:reference-motivation}
\end{figure}

    Observing the results for the first phase of the qualitative study, we find evidence that helps corroborate the findings of our quantitative models.
    Because we sampled developers with an extreme amount of ecosystem experience in our dataset, we can compare their contributions to developers who do not.
    We find that people with a normal amount of downstream experience make more content contributions ($F^*$, $p=0.001$, $odds=19.067$) and fewer meta contributions ($F^*$, $p<0.001$, $odds=0.041$).
    We see very similar results for people with low ecosystem experience who make more content contributions ($F^*$, $p<0.001$, $odds=21.250$) and found weaker evidence that they make fewer meta contributions ($F^*$, $p<0.001$, $odds=0.062$).
    Interestingly, we do not see the same pattern when comparing the contributions of the bottom two quartiles with the top two.
    Beyond the impact of intra-project experience reported in the previous paragraph, we only find weak evidence that people with low upstream experience make more content contributions ($F^*$, $p=0.067$, $odds=3.852$) and fewer meta contributions ($F^*$, $p=0.067$, $odds=0.260$).
    
    To identify whether people with ecosystem experience behave differently from those with intra-project experience we did a pair-wise comparison between the ecosystem strata and the intra-project strata.
    The results suggest that people with any type of ecosystem experience, in both the general case and newcomers, make fewer content contributions (for all: $F$, $p<0.001$, $odds<0.017$).
    In the general case, this translates to them making more meta contributions (for all: $F$, $p<0.001$, $odds>4.9$).
    For newcomers, we found evidence of this for upstream contributions ($F$, $p=0.048$, $odds=3.143$) and downstream contributions ($F$, $p=0.001$, $odds=5.762$), and weak evidence for general ecosystem experience ($F$, $p=0.078$, $odds=2.867$).
    Although we performed very similar tests to identify how frequently pull requests have clear ecosystem involvement, we only found weak evidence that the pull requests submitted by people who have made downstream contributions are less likely to explicitly reference the ecosystem compared to people with intra-project experience ($F$, $p=0.090$, $odds=0.307$).

    Proceeding to the statistical analysis of the reasons people reference ecosystem project in their pull requests, we find that references to the focal project (compared to references made by the focal project) are more commonly done when a change needs to be accommodated ($F^*$, $p=0.007$, $odds=3.088$) and less frequently to specify solutions ($F^*$, $p=0.031$, $odds=0.146$).
    We also find weak evidence that incoming references are less commonly done to describe a feature/bug/problem that is addressed ($F^*$, $p=0.052$, $odds=0.405$).
    
    Many outgoing references ($55.81\%$) were to GitHub projects owned by the same organization, clearly suggesting a more organized collaboration effort across various NPM projects that could explain our results.
    We found that people with a high community standing reference projects of the same organization much more frequently ($F^*$, $p=0.015$, $odds=24$).
    Similarly, we found weak evidence that newcomers with more downstream experience do this too ($F^*$, $p=0.056$, $odds=3.541$).
    Interestingly, newcomers with more ecosystem experience reference a project by the same organization much less often ($F^*$, $p=0.023$, $odds=0.031$).
    
    Notably, we do not find strong evidence that the reasons for referencing another GitHub project differ for the different types of ecosystem experience nor for the amount they have.
    However, there is some weak evidence that people with higher community standing reference other GitHub projects more frequently to motivate their changes compared to how often they do this to specify a solution ($\chi^2$, $p=0.063$, $odds=3.318$).
    In contrast, newcomers make references to motivate their changes less frequently compared to how often they specify a solution ($\chi^2$, $p=0.056$, $odds=0.434$).
    % Additionally, we see some weak evidence that newcomers are less likely to reference another project because they accommodate a change ($F^*$, $p=0.088$, $odds=0.414$).
}

% \subsection{Answering Questions}
\section{Discussion}
\label{sec:answering-questions}

The following sections discuss our results, answering the four research questions.

\subsection{Ecosystem-wide Contributions}
\label{sec:discussionecosystem-experience}

    The results of this study showed that a developer's ecosystem-wide contributions positively affect their pull request acceptance (shown in the \textit{ecosystem model} in Table~\ref{tab:coefficients-and-performance scores-general}).
    Although this notion might seem intuitive, it transcends previous claims as most previous studies emphasized the impact of intra-project technical experience \citep{tsay_influence_2014, iyer_effects_2021, rastogi_relationship_2018, zampetti_study_2019, soares_acceptance_2015, baysal_secret_2012, pinto_who_2018, lee_are_2017, gousios_exploratory_2014, khadke_predicting_2012, bosu_impact_2014}.
    Some studies considered the impact of ecosystem-wide experience, addressing its impact on pull request acceptance \citep{dey_effect_2020}, becoming a core contributor \citep{cheng_developer_2017}, and developers' joining behavior \citep{jergensen_onion_2011}.
    Our study transcends these studies twofold.

    Firstly, we consider non-coding contributions, namely participation in issue-tracking systems.
    This is relevant as recent studies indicated the prominence of non-coding contributions in OSS projects \citep{trinkenreich_hidden_2020, canovas_izquierdo_analysis_2021, geiger_labor_2021, trinkenreich_pots_2022}.
    Previous work already hinted towards the relevance of issue-tracking systems, as they are commonly used to discuss architectural knowledge \citep{soliman_exploratory_2021} and requirements \citep{perez-verdejo_requirements_2021}, and commonly complement pull requests \citep{alshara_pi-link_2023}.
    However, the relevance of non-coding contributions has not been empirically explored yet.
    We identify a positive relationship between the number of coding and non-coding contributions.
    Previous work suggested the importance of this factor concerning attaining \qt{committer} status in Apache projects \citep{gharehyazie_social_2013, gharehyazie_developer_2015}.
    Our study translates this result to pull-based development, finding that non-coding contributions matter when joining the project as a non-core contributor.

    Secondly, we separately addressed ecosystem-wide developer contributions in pull request decision-making, which \citet{dey_effect_2020} attempted to do as well.
    We separated intra-project contributions from ecosystem-wide contributions, eliminating the possibility of intra-project contributions confounding the results \citep{zhang_pull_2022, gousios_exploratory_2014, rastogi_relationship_2018, khadke_predicting_2012, zampetti_study_2019}.
    We successfully replicated the results by \citet{dey_effect_2020}, and nuance their claims by showing that, intra-project contributions have a 2\% larger impact on pull request decisions than ecosystem-wide contributions in the general case.
    
    \revOneAdd{Looking into the types of contributions made by these developers, we find strong evidence that people with ecosystem contributions generally make more meta contributions (e.g., dependency version bumps or changes to CI pipelines) compared to content contributions (e.g., bug fixes or new features).
    This pattern continues for contributors with a \textit{very} high amount of ecosystem contributions.
    This suggests that experience acquired in the ecosystem speeds up becoming an important member of other ecosystem projects.

    We created an early taxonomy of reasons to involve ecosystem projects in pull requests (see Table~\ref{tab:ref-motivations} and Figure~\ref{fig:reference-motivation}) by zooming into pull requests with clear ecosystem involvement.
    We find evidence that references to and from other ecosystem projects are generally done to motivate new changes or describe a solution.
    For example, when another pull request introduces a change that needs to be accommodated, or because the solution used in another project can also be applied in the focal project.
    Compared to references made by the focal pull request (i.e., from the pull request to other projects), references \textit{to} the focal pull request are less commonly done to specify similar solutions and are more commonly done when the proposed change needs to be accommodated in another project.
    In other words, because incoming and outgoing references are for different reasons, we find early evidence that pull requests involve the ecosystem differently from how the ecosystem involves pull requests.
    }

    \begin{insight}{1}~\textit{How do a developer's {coding and non-coding contributions in the ecosystem affect the acceptance of their pull request submissions?}}~\\~\\
        Pull request acceptance decisions depend on more than merely intra-project factors and code-oriented contributions.
        Non-coding participation through participation in issue-tracking systems and code reviews positively affects pull request decisions.
        % At an intra-project and ecosystem level, non-coding participation in issue-tracking systems and code reviews positively affects pull request decisions.
        \revOneAdd{Ecosystem experience also affects the tasks performed, shifting from content contributions (features, bugs, etc.) to meta contributions (CI pipelines, dependency maintenance, etc.), suggesting ecosystem-wide contributors take prominent roles in projects.
        We identified 3 overarching and 10 specific reasons to reference ecosystem projects.
        Pull requests with ecosystem involvement commonly involve other projects to motivate proposed changes or to reference similar solutions.
        We found early evidence that ecosystem references made by pull requests are for different reasons than references made to pull requests.}
    \end{insight}

\subsection{Contributions in Dependent Projects}
\label{sec:dependency-ecosystem-expereince}

    Although understanding the impact of general ecosystem contributions is important, ecosystems are special because projects often depend on each other \citep{decan_empirical_2019}.
    Dependencies are defined in two directions: upstream and downstream, such that downstream projects implement the functionality of an upstream project.
    This means that projects have some technical relevance to each other, as experience implementing a package could help improve it \citep{maeprasart_understanding_2023, palyart_study_2018, rehman_newcomer_2022, shah_motivation_2006, subramanian_analyzing_2022} and building a package could help implement it \citep{bogart_how_2016, bogart_when_2021}.
    \revOneAdd{Therefore, differentiating between these types of experience allows us to see the impact of \textit{where} in the ecosystem you contributed rather than only addressing \textit{if} you have (like we did in \rqn{1}).}
    
    The results in the \textit{dependency model} in Table~\ref{tab:coefficients-and-performance scores-general} suggest that contributions in upstream and downstream projects positively affect pull request decision outcomes.
    This is consistent with previous findings \citep{dey_effect_2020} identifying the positive impact of downstream contributions.
    The comparison of upstream and downstream contributions suggests that downstream contributions have a 3\% larger impact than upstream contributions.
    This could be partially explained through the different roles that upstream and downstream developers have in dependency networks.
    Although downstream developers are paramount when fixing issues and building new features \citep{palyart_study_2018, shah_motivation_2006, valiev_ecosystem-level_2018, geiger_labor_2021}, and in some cases even receive priority with problems \citep{geiger_labor_2021}, limited evidence has been found regarding the benefits of upstream developers.
    Although \citet{bogart_how_2016,bogart_when_2021} find that upstream developers occasionally help downstream projects integrate breaking API changes, they explicitly conclude this is uncommon in the NPM ecosystem.

    To contextualize the relevance of dependency contributions, a comparison has been made with non-dependency ecosystem contributions.
    These are contributions made in neither upstream nor downstream projects.
    The results suggest that the impact of non-dependency contributions is $11.8\%$ and $8.7\%$ more important than upstream and downstream contributions, respectively.
    Although dependency contributions are relevant, the added specific knowledge of a technically related package does not outweigh the relevance of the general contributions of the ecosystem’s common technology.
    Alternatively, this could mean that the tasks picked up by this group are more complex.

    \revOneAdd{
        Similar to the results for \rqn{1}, we see that people (newcomers and non-newcomers) with downstream or upstream contributions make less content and more meta contributions.
        Although we see a trend that people with an increasing amount of downstream contributions make increasingly more meta and fewer content contributions, we cannot conclusively say this about upstream contributions.
        % This could relate to the difference in coefficients measured in the regression models.
        Because the shift in tasks performed is less clear as people made more upstream contributions, it reflects in the acceptance rate of their pull requests.
        Consequently, it highlights a potential difference in the applicability of upstream versus downstream experience.
    }
    
    Previous work vouched for the benefits of dependencies that are congruent with the developers working on the involved projects, finding that it can positively affect the longevity of projects \citep{valiev_ecosystem-level_2018, wattanakriengkrai_giving_2023} and that contributors tend to join upstream and downstream projects \citep{muller_role_2023}.
    Regardless, dependency networks are not always congruent with the developers working on projects \citep{syeed_socio-technical_2014}.
    Our results complement these studies in two fashions: 1) by showing its positive impact on a developer’s technical success, as indicated by their pull request acceptance, and 2) that, collaborating with dependent projects has clear benefits, it only partially explains developer success as contributions made in non-dependent projects have a stronger impact on pull request decisions.
    
    \begin{insight}{2}~\textit{How do a developer's contributions to {upstream and downstream projects} affect the acceptance of their pull request submissions?}~\\~\\
        Our results suggest that contributions in downstream and upstream projects positively affect pull request decisions.
        \revOneAdd{We see that contributors with upstream/downstream contributions make more meta and fewer content contributions than developers with intra-project experience.}
        In general, downstream contributions are more important than upstream contributions.
        \revOneAdd{To some extent, this coincides with the tasks they perform such that people with increasing downstream experience submit more meta changes, while we cannot see the same pattern for contributors with upstream experience.}
        % This could indicate the general usefulness of upstream compared to downstream experience.}
        This highlights the importance of dependency-specific knowledge in cross-project collaboration.
        However, \revOneAdd{our regression models suggest that }contributions made in upstream/downstream projects are less impactful than contributions made in non-dependent projects.
    \end{insight}

\subsection{Direct Collaboration and Ecosystem-wide Standing}
\label{sec:discussion-collaboration}

    Open-source software development is commonly described as a socio-technical process \citep{forte_defining_2013} where people are judged on their technical merit \citep{wermke_committed_2022, steinmacher_social_2015}, but might be trusted earlier because of their social connections \citep{soto_analyzing_2017, yu_determinants_2016, zhang_pull_2022, iyer_effects_2021, tsay_influence_2014}.
    \revOneAdd{While the answers to \rqn{1,2} highlighted the impact of the developers' contributions, they give limited insights into the \textit{collaborative} efforts of open-source communities.}
    Prior research showed that social connections can positively affect project longevity \citep{hahn_impact_2006, wang_survival_2012, fang_matching_2023}, the number of pull requests and forks a project receives \citep{celinska_coding_2018}, and project downloads \citep{mendez-duron_returns_2009, fershtman_direct_2011, peng_network_2013}.
    Further, people tend to join projects they are socially connected to \citep{casalnuovo_developer_2015, hahn_emergence_2008, de_souza_social_2016}.

    Although these studies give insights into the likelihood of joining and \textit{successfully} joining a project, they limitedly explain the impact of social connectedness while submitting a pull request.
    Several studies have addressed this at an intra-project level, identifying a relationship with prolonged project participation \citep{qiu_going_2019}, pull request acceptance \citep{gharehyazie_social_2013, gharehyazie_developer_2015, bosu_impact_2014}, task performance \citep{carillo_what_2017}, and in some cases productivity \citep{zoller_topology_2020}, at the risk of higher bug proneness \citep{chen_empirical_2024}.
    Although this gives a good overview of the impact of social connectedness inside a project, it limitedly explains the impact at an ecosystem scale.

    Studies that transcend a project attempted to fill this gap using GitHub's follower feature, finding that following your pull request integrator \citep{soto_analyzing_2017, yu_determinants_2016, zhang_pull_2022}, and being well-connected in general \citep{iyer_effects_2021, tsay_influence_2014, zhang_pull_2022} are positively related to pull request acceptance.
    Although these findings highlight the impact of social network features in social coding platforms like GitHub, they can only limitedly represent the professional network.
    This is because the follower feature can be used by anyone for any reason, without much regard for their professional relationship.
    We constructed a professional social network, extracting connections between collaborators using the pull requests and issues they were involved in, ensuring two connected developers have collaborated in the past.
    Participation in the ecosystem is measured in two fashions, \textit{direct collaboration} and \textit{ecosystem-wide community standing}, to identify the impact of direct proximity to your pull request integrator and your general connectedness in the ecosystem on pull request decisions, respectively.
    The results in the \textit{collaborative model} in Table~\ref{tab:coefficients-and-performance scores-general} suggest that both factors positively affect pull request acceptance.
    
    It stands out that direct collaboration has a 12\% larger influence on pull request decisions than intra-project contributions.
    This underlines the importance of professional connectedness to contribution success and can be explained through higher pre-established trust between them.
    Regardless of whether the developer might still need to learn the project's specifics, this prior connection might be sufficient to vouch for their competencies.
    Alternatively, because the developers have collaborated already, they might be more familiar with each other's manner of working (e.g., coding style, or manner of explaining things), for which it could be easier to achieve a fruitful discussion and consensus on implementation details.
    \revOneAdd{This feature was unimportant in the predictive model, explaining only $0.6\%$ of the information.
    This is not surprising because, at an ecosystem level, collaboration between a pull request submitter and a pull request integrator is relatively rare, as this occurred in only $10\%$ of the cases.
    Testing for this subset of data significantly increased its importance, becoming the seventh most important feature, just below ecosystem-wide community standing.}
    
    We also find that ecosystem-wide community standing has a positive effect on pull request decisions; however, $3.5\%$ less than intra-project contributions and $12\%$ less than direct collaboration.
    This can be explained because the metric is less tangible --- as it measures if someone has collaborated with many or experienced developers.
    This result can be explained similarly to \citet{wang_survival_2012}, who studied the external network of a project; i.e., the number of developers a project is connected to because one of its contributors also collaborates in other projects.
    \citeauthor{wang_survival_2012} reasons that having an increased number of connections increases the likelihood that a developer interacts with and consequently acquires high-quality resources or deeper knowledge of the system.
    In our case, because a developer collaborated with many or experienced others in the ecosystem, they might have learned the specific knowledge necessary to produce a high-quality pull request, which is more likely to be accepted.

    \revOneAdd{Like \rqn{1,2}, we can see that the pull requests made by people with ecosystem collaborations commonly contain more meta and fewer content contributions.
    We have already seen that over half of the references we analyzed are to other projects owned by the same organization as the focal project (e.g., CKEditor, CanJS, or Microsoft).
    Clearly, this could explain our results, as organizations are collaborative initiatives by definition.
    Consequently, further analysis showed that people with high community standing reference projects from the same organization more frequently, suggesting our metric can capture these communities.
    Interestingly, we find no evidence that the same is true for developers who collaborated directly with their pull request integrator, suggesting a less organized collaboration.
    }
    
    \begin{insight}{3}~\textit{How do a developer's {ecosystem-wide community standing and direct collaboration with their integrator in the ecosystem} affect the acceptance of their pull request submissions?}~\\~\\
        Our results suggest that direct collaboration with your integrator and your ecosystem-wide community standing positively relate to pull request acceptance.
        \revOneAdd{We see that both groups generally make fewer content contributions than meta contributions, suggesting greater seniority.}
        This underlines the importance of professional connections at an ecosystem level.
        \revOneAdd{Looking deeper, we find that references in pull requests submitted by developers with high community standing reference projects owned by the same organization as the focal project much more often.
        This highlights the importance of organized collaborative efforts in research on pull requests.}
        We further measure that direct collaboration with your integrator is more important than intra-project contributions, highlighting the importance of social factors as a complement to project-specific knowledge.
    \end{insight}

\subsection{Project Newcomers}
\label{sec:first-time-contributors}

    \revOneAdd{Studying software ecosystems yields the unique opportunity to study how developers move through the ecosystem and test the impact of previous collaborations and contributions on that movement.
    The results of \rqn{1-3} provide an overview of developers' ecosystem participation and its general impact on their pull request submissions.
    We specifically address when users join a project for the first time by studying the impact of ecosystem participation on newcomers' pull request submissions.}
    Project newcomers stand at the foundation of OSS projects, as the threshold of joining and leaving is low \citep{forte_defining_2013}, and having more newcomers can positively affect project longevity \citep{qiu_going_2019}.
    
    Regardless of their fundamental role in OSS projects, various studies have suggested that large projects are more reluctant to let new people in \citep{palyart_study_2018}, and that pull requests submitted by newcomers are more frequently refused \citep{kovalenko_code_2018, lee_are_2017, soares_acceptance_2015, soares_rejection_2015} --- a finding that was supported in this study, as their pull requests were $12.9\%$ more commonly refused (see Table~\ref{tab:coefficients-and-performance scores-general}).

    Various socio-technical barriers have been identified that prevent newcomers from successfully onboarding in open-source projects \citep{steinmacher_social_2015, steinmacher_let_2019, steinmacher_overcoming_2019, geiger_labor_2021}.
    An example of this is the perceived lack of expertise, which was identified in the works of \citet{wermke_committed_2022} and \citet{rehman_newcomer_2022}, listing it as a means to gain trust (through meritocracy) and a hurdle that prevents them from successfully onboarding, respectively.
    This barrier is not only present in open-source projects, as \citet{rastogi_ramp-up_2015} identified the same phenomenon in closed-source projects.

    Various socialization factors are listed as a barrier for newcomers \citep{steinmacher_social_2015, steinmacher_let_2019, steinmacher_overcoming_2019}, like low-quality discourse or low responsiveness.
    The work of \citet{gharehyazie_social_2013, gharehyazie_developer_2015} and \citet{carillo_what_2017} zoomed in on this topic, both studying the impact of intra-project socialization on onboarding in open-source projects.
    \citet{gharehyazie_social_2013, gharehyazie_developer_2015} identified a positive relationship between socialization in the project's mailing list and becoming a committer, and \citet{carillo_what_2017} that it positively relates to newcomers' task performance.

    Each of these studies vouches for the importance of developer experience and collaboration at an intra-project level, showing positive effects on newcomer onboarding.
    This leaves a gap in current literature that explores their impact at an ecosystem scale.
    Our work operationalizes a hypothesis by \citet{rastogi_how_2021}, proposing that \qt{unit changes from new contributors originating in the same ecosystem are more likely to be accepted.}

    % Compares all variables with intra-project experience.
    Our results in Table~\ref{tab:coefficients-and-performance scores-ftc} suggest that each ecosystem-wide contribution and collaboration factors positively affect the outcomes of newcomers' pull request decisions.
    It stands out that the measured impact of all variables exceeds that of intra-project contributions for newcomers (through non-coding contributions like posting an issue) between $8\%$ and $33\%$.
    This is substantially different from non-newcomers and the two groups combined, as we observe the inverse such that intra-project contributions yield between $0.5\%$ and $13\%$ greater impact than the ecosystem-wide contribution and collaboration metrics.
    The only exception is direct collaboration between the pull request submitter and integrator, which has a $13\%$ larger impact than intra-project contributions in the general case.
    However, this is notably smaller than the $33\%$ increase measured for newcomers.
    The predictive models for newcomers' pull requests also depend more strongly on these variables.
    
    % More detailed results of dependencies
    We further observe that all three ecosystem contribution types (i.e., upstream, downstream, and non-dependency contributions) affect pull request decisions $8$--$11\%$ more than intra-project contributions.
    Additionally, upstream contributions exceed downstream contributions by $2.3\%$, which is the opposite for non-newcomers.
    A potential explanation would be that experience acquired through upstream contributions translates better to downstream projects compared to the inverse.
    \revOneAdd{However, like we saw in \rqn{1,2}, newcomers with ecosystem contributions perform different tasks as well.
    Regardless of being new, they make fewer content contributions than developers with intra-project experience, and newcomers with upstream or downstream contributions make more meta contributions.
    Contrary to expectation, this difference is much greater for developers with upstream contributions than those with downstream contributions.
    Therefore, although the difference in skill transferability could still be true, a more concrete explanation is that they perform different tasks, which benefits newcomers with upstream experience in pull request acceptance.}
    % A potential explanation for this phenomenon is that experience acquired through upstream contributions translates better to downstream projects compared to the inverse.
    For example, implementing a new feature based on an updated API largely requires knowledge of the API and local knowledge of where it is implemented, whereas adding a new feature to an upstream project requires an in-depth understanding of the other project.
    This could suggest that having contributed to a project that the focal project is aware of (a project they have already integrated) might yield an increased level of trust initially.
    However, as the developer continues to contribute, picking up more difficult tasks, its impact decreases.
    Similarly, changes proposed to upstream projects might be more complicated.

    % More detailed results of collaboration
    Expanding on the results of \rqn{3}, a substantial difference can be observed in the impact of the newcomers' ecosystem-wide community standing and their prior direct collaborations with their pull request integrator, increasing the likelihood of pull request acceptance with $15\%$ and $33\%$, respectively.
    Interestingly, whereas the impact of community standing is $3\%$ lower than intra-project experience for non-newcomers, direct collaboration with their pull request integrator remains a stronger indicator of pull request acceptance, outweighing intra-project contributions by $14\%$.
    This highlights the importance of ecosystem-wide collaboration for non-newcomers.
    \revOneAdd{Again, similar to \rqn{3}, newcomers with ecosystem collaborations perform notably fewer content contributions.
    However, we could not find evidence that they perform more meta contributions.}

    % Prediction
    Interestingly, \revOneAdd{we see that the feature importance of \qt{PR submitter is a newcomer} explains only $1\%$ of the prediction.
    This could be due to the random effects across projects which could potentially influence random forest outcomes.
    However, another explanation could be that predicting the outcome of pull requests submitted by newcomers remains difficult, as Table~\ref{tab:model-performance} only reports an F1 score of $0.83$ even though the impact of ecosystem-wide variables shown in Table~\ref{tab:coefficients-and-performance scores-ftc} is high.}
    % However, although the impact of ecosystem-wide variables shown in Table~\ref{tab:coefficients-and-performance scores-ftc} is high, predicting the outcome of pull requests submitted by newcomers remains difficult, as Table~\ref{tab:model-performance} only reports an F1 score of $0.83$.
    This score is noticeably lower than the classification performance of non-newcomers, having an F1 score of $0.96$.
    Including ecosystem-wide factors increased the classification performance by $2.3\%$, which is noticeably higher than the impact for non-newcomers, increasing only $0.2\%$.
    We conclude that ecosystem-wide collaboration and contribution factors are a meaningful addition to models predicting the outcome of pull requests submitted by newcomers regardless of the generally lower prediction performance compared to pull requests submitted by non-newcomers.

    Our results support the hypothesis presented by \citet{rastogi_how_2021}, showing that ecosystem-wide experience positively affects the success of newcomers in software projects, as the results suggest the positive impact of 1) prior domain knowledge and 2) a connection built with the development team before proposing a change.
    \revOneAdd{We observe that newcomers with ecosystem participation make different contributions too.}
    Therefore, although participating in intra-project activities might help shed light on implementation considerations that otherwise might have been overlooked, explain the community development guidelines, or act as a means to establish trust with other members, similar socio-technical factors exist at an ecosystem level, such that the competencies acquired in one project naturally translate themselves to another in the same ecosystem.
    This is especially true for newcomers as they have limited knowledge of the focal project, inherently depending on everything outside it.
    
    \begin{insight}{4}~\textit{How do {a project newcomer's contributions and collaborations in the ecosystem} affect the acceptance of their pull request submissions?}~\\~\\
        Our results suggest that ecosystem-wide contributions in upstream, downstream and non-dependency projects positively affect newcomers' pull request decisions and are a stronger indicator of pull request acceptance than prior non-coding participation inside the project.
        \revOneAdd{Newcomers with prior ecosystem participation make fewer content contributions and in some cases more meta contributions.}
        \revOneAdd{Our results nuance the difference between upstream and downstream developers.
        Although pull requests submitted by newcomers with upstream contributions are more commonly accepted, they also perform different tasks than people with intra-project experience.}
        Direct collaboration with your pull request integrator in the ecosystem and a newcomer's ecosystem-wide community standing strongly and positively affects their pull request decisions.
        This underlines the existence and importance of ecosystem-wide socio-technical factors in open-source software development, which can translate across projects in the same ecosystem.
    \end{insight}
    
    \section{Implications}
\label{sec:implications}
    
Our work is relevant to research and practice as the ecosystem perspective is becoming more critical in software engineering.
This is due to the increased emphasis on software supply chains and the intricate dependencies between the many systems of today.
It is paramount for organizations and individuals to understand how people engage across the entire ecosystem, not only in projects.

Our results show that upstream/downstream/non-dependency ecosystem-wide experience and collaboration are important factors in pull request decisions, especially for newcomers.
\revOneAdd{We find early evidence that ecosystem developers perform different tasks in projects compared to developers who commonly contribute to it.
This could suggest that ecosystem developers provide different value to projects.

We constructed a preliminary taxonomy of how projects involve other ecosystem projects in their development process.
Among others, we see that events in other projects commonly motivate a change (e.g., because an update must be accommodated) or that other projects implemented related solutions.}

Our results highlight the importance of non-coding contributions, like discussions in issue-tracking systems.
Although these address notably different goals in software projects than coding activities, they strengthen skills developers need to collaborate in the broader ecosystem.
Organizations can leverage these results to identify the group of external developers who can contribute to their projects.
The NPM community can learn from other OSS ecosystems, as previous work has suggested that collaboration between projects is less common in the NPM ecosystem \citep{bogart_how_2016, bogart_when_2021}.
They could strengthen cross-project cooperation and reciprocity, allowing projects to grow as an ecosystem rather than separate entities in a shared environment.

Contributors (including project newcomers) can use these socio-technical relationships to navigate the ecosystem.
For example, it might be beneficial to join a project in the ecosystem's periphery and move to more central projects over time.
Our results suggest that pull requests integrated by a professional acquaintance are substantially more likely to be accepted.
This suggests a valuable strategy to navigate an ecosystem: to join the projects your professional peers already partake in.
Combined with recommended intra-project joining behaviors (e.g., \citet{steinmacher_let_2019}, \citet{gharehyazie_social_2013, gharehyazie_developer_2015}, and \cite{carillo_what_2017}), like participating in project discussions before proposing a change and submitting a pull request, could further speed up the project onboarding process.
This could greatly benefit people having difficulty joining the community, whereas they would improve its productivity, diversity, inclusion, and overall well-being.

The ecosystem perspective affects research discussing barriers that newcomers experience when joining projects \citep{steinmacher_social_2015, steinmacher_almost_2018, steinmacher_let_2019, steinmacher_overcoming_2019, wermke_committed_2022, rehman_newcomer_2022, rastogi_ramp-up_2015, geiger_labor_2021} because these barriers are observably lowered for experienced newcomers, as indicated by the results of our work.
\revOneAdd{We find early evidence that collaboration across projects owned by the same organization might be better than across projects that are not.
Therefore, future research should explicitly consider movements between projects owned by the same organization/users versus those owned by different ones.}
Future research should address the extent to which developers migrate across open-source ecosystems, identify the best practices and common downfalls of this process, and explore the benefits and drawbacks of these migrations on open-source projects.
For example, continued collaboration between developers inside a project (i.e., when they have worked together a lot in that project) has been linked to bug proneness by \citet{chen_empirical_2024}.
This could translate to an ecosystem level, such that continued collaboration in one project introduces risks in another project they work on together.
Beyond yielding insights into how ecosystem-wide collaboration changes software projects as single entities \citep{rastogi_how_2021}, it also teaches us how software ecosystems change as a whole.

\section{Threats to Validity}
\label{sec:threats-to-validity}

\subsection{Internal Validity}
\label{sec:internal-validity}

    The collection of the projects studied in this work was largely based on their popularity in the NPM ecosystem.
    Although this set was complemented by dependency projects, all of which are less popular, it does not ensure that the general representation of NPM projects improved.
    Additionally, this method introduced a stark skewness in the distribution of pull requests across projects, such that $2\%$ of the projects contained $49\%$ of the pull requests.
    By randomly sampling these pull requests, we reduced the impact of this inequality.
    \revOneAdd{Beyond project size, our results could be similarly biased towards extremely active contributors and organizations who own many projects.
    However, through exploratory analysis where we excluded these groups, we could not identify their impact, suggesting they minimally affected our results.}

    People commonly use multiple aliases in platforms like GitHub \citep{wiese_who_2016}, which can substantially affect the structure of social networks.
    Although various solutions exist to merge aliases \citep{amreen_alfaa_2020, vasilescu_data_2015}, none were applied due to stringent data requirements or the large amount of required manual labor to accurately apply these at an ecosystem scale.
    Furthermore, although this study intended to only include activities performed by humans, filtering out anything performed by bots using previous work \citep{dey_detecting_2020, dey_dataset_2020, golzadeh_ground-truth_2021, golzadeh_ground-truth_2020} as well as manual analysis of users with over 400 pull requests, we cannot assure that all bots were removed.
    \revOneAdd{This specifically regards bots that use user accounts, and accounts used by bots \textit{and} real users.}
    
    The dataset created by \citet{katz_librariesio_2020} is the foundation of this study because of the large amount of archival data it contains.
    Our study inherits the flaws in this dataset, one of which is the accuracy of the identified dependencies.
    The data represents a snapshot without regard for changes happening over time.
    Therefore, although the listed dependencies were correct at some point, they might not accurately represent the project's real dependencies over time \citep{decan_empirical_2019}.
    Although sometimes projects do not list their dependencies, this did not impact our results as this was only the case in 2\% of the projects in our dataset.

    \revOneAdd{To strengthen the internal validity of our work, we performed an exploratory qualitative analysis in addition to our quantitative study.
    Although this gives early insights into why we observe the positive impact of ecosystem participation on pull request acceptance and an early taxonomy of motivations to include the ecosystem in pull request discussions, our qualitative results must be interpreted carefully and verified in future work.
    This specifically concerns the representativeness of our sample, which emphasizes references made by contributors with ecosystem experience and is, therefore, not representative of the NPM as a whole.}
    
\subsection{Construct Validity}
\label{sec:construct-validity}

    We employed two social network analysis metrics to measure direct collaboration and ecosystem-wide community standing: \textit{link strength} and \textit{node centrality}.
    Although the latter can be measured in many ways, we applied \textit{second-order degree centrality} because this has much lower computational complexity than commonplace metrics, like Eigenvector centrality \citep{newman_measures_2018}.
    \revOneAdd{This metric was inspired by previous work who used it to influential people \citep{brodka_analysis_2012, chen_identifying_2012} and trust \citep{stickgold_trust_2013} in social networks, and core developers \citep{jergensen_onion_2011}.}
    Although our metric is coarser, it is representative as it captures the impact of a collaborator's neighbors, which is essential when determining the node's centrality \citep{newman_measures_2018}.
    \revOneAdd{Our qualitative results further support this, as its results suggested that the projects referenced in pull requests submitted with high centrality reference projects owned by the same organization more often, indicating an organized \textit{collaborative} effort.}
    
    Further, this study identified a relationship between ecosystem-wide variables and pull request decisions.
    To ensure our results reflect reality, several control variables recommended by \citet{zhang_pull_2022} were included, allowing us to contextualize our results.
    We omitted one recommended variable regarding continuous integration because it is difficult to collect this data.
    However, after studying the results of \citet{zhang_pull_2022} in detail, we have no reason to believe this affects our results.
    
    \revOneAdd{We include two models, mixed-effects logistic regression and random forest, explaining our metrics' impact and predictive strength.
    Although we specifically chose a \textit{mixed-effects} model for the former, which can account for differences across projects, we do not account for this in the random forest, even though its reported feature importance can be affected by them.
    However, we find very limited evidence of this in our context because the results reported by random forest align with those reported by mixed-effects logistic regression, such that features with high impacts in the mixed-effects logistic regression models also have high feature importance.
    The only exceptions here are \qt{PR submitter is a newcomer} and \qt{direct collaboration.}
    However, we find that the limited importance of direct collaboration is due to its sparsity.
    The limited importance of being a newcomer is likely because pull requests submitted by newcomers are generally harder to predict.}

\subsection{External Validity}
\label{sec:external-validity}

    This work addresses NPM, for which our conclusions might not represent other ecosystems \revOneAdd{because the type of ecosystem and its common technologies (e.g., the language) may influence how people contribute.}
    Regardless, we think these results could translate to other packaging ecosystems (e.g., PyPI or Maven) as they are similar to NPM.
    \revOneAdd{Specifically, because projects in these ecosystems are known for reusing functionality of other projects, like they do in NPM.
    However, future work should verify this.}
    Further, \citet{valiev_ecosystem-level_2018} identified the relevance of cascading dependencies \citep{geiger_labor_2021}.
    We did not consider this because little is known about it in the literature and because of the complexity it would add to our analysis.
    Future work could address this.
    Consequently, our results are limited to only direct dependencies between projects.

\section{Conclusion}
\label{sec:conclusion}

Our study objective was to explore the impact of ecosystem-wide experience and collaboration on pull request acceptance in NPM.
It is crucial to understand this relationship for two reasons. 
One, software projects are no longer developed in isolation \citep{decan_empirical_2019}, \revOneAdd{posing a significant risk to project developers because bugs can propagate from one system to another \citep{ma_impact_2020}.}
Therefore, to understand software development better \revOneAdd{and stimulate cross-project collaboration,} we must understand the software ecosystem it comprises.
Two, developers and projects seek meaningful ways to engage in open-source software development to attract new contributors, \revOneAdd{starting new collaborations.}
\revOneAdd{Therefore, understanding how contributors move between projects in your ecosystem is paramount to sustaining a healthy group of contributors.}

Our study explores four topics: 1) general ecosystem-wide coding and non-coding experience, 2) experience in upstream/downstream projects, 3) direct collaboration and community standing in ecosystems, and 4) the impact of ecosystem-wide experience and collaboration on pull requests submitted by project newcomers.
We investigate these relationships on 1.8 million pull requests and 2.1 million issues from 20,052 projects of the NPM ecosystem.

The results show that ecosystem-wide experience and collaboration positively affect pull request acceptance decisions.
In general, ecosystem-wide metrics do not exceed the importance of developers' intra-project experience, except for prior collaboration in the ecosystem between the pull request integrator and submitter. 
The opposite is true for project newcomers, as the impact of all ecosystem-wide variables exceeded that of intra-project contributions like participating in discussions in the issue-tracking system.

\revOneAdd{Our exploratory manual analysis on 538 pull requests identified 1)~why we see this positive impact and 2)~how developers involve the ecosystem in their pull request discussions.
We see that developers (including newcomers) with ecosystem experience generally make fewer content contributions (i.e., new features, bug fixes, etc.) and, under certain circumstances, make more meta contributions (e.g., changes to the CI pipeline or dependency version bumps).
We found 111 pull requests with clear ecosystem involvement, allowing us to develop an early taxonomy of 3 overarching and 10 specific reasons to reference ecosystem projects.
This is most commonly done to motivate a new change or to reference a solution in another project.}

The results show that ecosystem-wide experience and collaboration factors complement previously known factors when predicting the outcomes of pull requests.
Our metrics increased their classification performance between $0.2\%$ and $2.3\%$, yielding an overall F1 score of $0.83$ for pull requests submitted by project newcomers, $0.96$ for non-newcomers, and an overall F1 score of $0.92$, \revOneAdd{and highlights the increased challenge of predicting the outcome of pull requests submitted by newcomers.}

\revOneAdd{These results corroborate the relevance of ecosystem-wide participation, indicating the importance of \textit{if} you do so, \textit{where} you do it, and with \textit{whom} you did.
Future work could take various directions.
For example, studying other ecosystems (packaging ecosystems, like PyPI or Maven, and thematic ones, like OpenStack or NumFocus), the impact of organizations on cross-project collaboration, the difference in contributions made by upstream and downstream contributors, or the resolution of problems that span multiple projects.}

    \section*{Conflict of Interest}
        The authors have no competing interests to declare that are relevant to the content of this article.

    \section*{Data Availability Statement}
        The replication package of this paper can be found on Zenodo \citep{meijer_2024_dataset} (or click the following link: \href{https://zenodo.org/records/13286685?preview=1&token=eyJhbGciOiJIUzUxMiJ9.eyJpZCI6ImQ3NTgxZGJmLWJiZDktNGRjNy05ZDRiLTdiYjQxMDhlNThjNSIsImRhdGEiOnt9LCJyYW5kb20iOiI2OWMwOTYyYzc4YWRlMDE1YzNjZjViYTRlZWZkYmY0ZCJ9.pYOT73rp0pw9pjaal3bYZ98TL-gnr6BplrcO_YeQopFiHo2yEVfNhYgFwkbRLwexnV3jdGuMss7XNS1t3f0Zyw}{zenodo.org/records/13286685}).

    \begin{acknowledgements}
        \revOneAdd{We thank the Center of Information Technology (CIT) at the University of Groningen for providing the computing resources necessary to carry out this study.}
        We thank D{\'a}niel Varr{\'o} \revOneAdd{and Kristian Sandahl for their} insightful feedback and suggestions. 
        Their expertise and constructive comments have been greatly appreciated and were crucial in shaping the final version of this paper.
    \end{acknowledgements}

% Toggle the following to 1 / 0 based on whether the document is WIP or being prepared for submission.
% \def\bibmode{1}

% \if 1\bibmode
%     % Forces some rules onto the bibliography.
%     \begin{filecontents}{natbibconfig.bib}
%         @IEEEtranBSTCTL{BibliographyConfig,
%           CTLuse_forced_etal       = "yes",
%           CTLmax_names_forced_etal = "2",
%           CTLnames_show_etal       = "1",
%           CTLdash_repeated_names   = "no"
%         }
%     \end{filecontents}
%     \bstctlcite{BibliographyConfig}
% \fi

\bibliographystyle{spbasic}

% See details for "bibmode" above.
% \if 0\bibmode
%     \bibliography{bibliography/zotero,bibliography/other_references}
% \else
% \bibliography{bibliography/cleaned_references,bibliography/other_references}
\bibliography{doi2bib_references}

\begin{thebibliography}{109}
\providecommand{\natexlab}[1]{#1}
\providecommand{\url}[1]{{#1}}
\providecommand{\urlprefix}{URL }
\expandafter\ifx\csname urlstyle\endcsname\relax
  \providecommand{\doi}[1]{DOI~\discretionary{}{}{}#1}\else
  \providecommand{\doi}{DOI~\discretionary{}{}{}\begingroup
  \urlstyle{rm}\Url}\fi
\providecommand{\eprint}[2][]{\url{#2}}

\bibitem[{Alshara et~al.(2023)Alshara, Shatnawi, Eyal-Salman, Seriai, and
  Shatnawi}]{alshara_pi-link_2023}
Alshara Z, Shatnawi A, Eyal-Salman H, Seriai AD, Shatnawi M (2023) {PI}-link: A
  ground-truth dataset of links between pull-requests and issues in {GitHub}.
  IEEE Access 11:697–710, \doi{10.1109/access.2022.3232982},
  \urlprefix\url{http://dx.doi.org/10.1109/ACCESS.2022.3232982}

\bibitem[{Amreen et~al.(2020)Amreen, Mockus, Zaretzki, Bogart, and
  Zhang}]{amreen_alfaa_2020}
Amreen S, Mockus A, Zaretzki R, Bogart C, Zhang Y (2020) {ALFAA}: Active
  learning fingerprint based anti-aliasing for correcting developer identity
  errors in version control systems. Empirical Software Engineering
  25(2):1136–1167, \doi{10.1007/s10664-019-09786-7},
  \urlprefix\url{http://dx.doi.org/10.1007/s10664-019-09786-7}

\bibitem[{Badampudi et~al.(2023)Badampudi, Unterkalmsteiner, and
  Britto}]{badampudi_modern_2023}
Badampudi D, Unterkalmsteiner M, Britto R (2023) Modern code reviews—survey
  of literature and practice. ACM Transactions on Software Engineering and
  Methodology 32(4):1–61, \doi{10.1145/3585004},
  \urlprefix\url{http://dx.doi.org/10.1145/3585004}

\bibitem[{Barab{\'a}si and Albert(1999)}]{barabasi_emergence_1999}
Barab{\'a}si AL, Albert R (1999) Emergence of scaling in random networks.
  Science 286(5439):509–512, \doi{10.1126/science.286.5439.509},
  \urlprefix\url{http://dx.doi.org/10.1126/science.286.5439.509}

\bibitem[{Baysal et~al.(2012)Baysal, Kononenko, Holmes, and
  Godfrey}]{baysal_secret_2012}
Baysal O, Kononenko O, Holmes R, Godfrey MW (2012) The secret life of patches:
  A {Firefox} case study. In: 2012 19th Working Conference on Reverse
  Engineering, IEEE, \doi{10.1109/wcre.2012.54},
  \urlprefix\url{http://dx.doi.org/10.1109/WCRE.2012.54}

\bibitem[{Baysal et~al.(2015)Baysal, Kononenko, Holmes, and
  Godfrey}]{baysal_investigating_2016}
Baysal O, Kononenko O, Holmes R, Godfrey MW (2015) Investigating technical and
  non-technical factors influencing modern code review. Empirical Software
  Engineering 21(3):932–959, \doi{10.1007/s10664-015-9366-8},
  \urlprefix\url{http://dx.doi.org/10.1007/s10664-015-9366-8}

\bibitem[{Bogart et~al.(2016)Bogart, K\"{a}stner, Herbsleb, and
  Thung}]{bogart_how_2016}
Bogart C, K\"{a}stner C, Herbsleb J, Thung F (2016) How to break an {API}: cost
  negotiation and community values in three software ecosystems. In:
  Proceedings of the 2016 24th ACM SIGSOFT International Symposium on
  Foundations of Software Engineering, ACM, FSE’16, p 109–120,
  \doi{10.1145/2950290.2950325},
  \urlprefix\url{http://dx.doi.org/10.1145/2950290.2950325}

\bibitem[{Bogart et~al.(2021)Bogart, K\"{a}stner, Herbsleb, and
  Thung}]{bogart_when_2021}
Bogart C, K\"{a}stner C, Herbsleb J, Thung F (2021) When and how to make
  breaking changes: Policies and practices in 18 open source software
  ecosystems. ACM Transactions on Software Engineering and Methodology
  30(4):1–56, \doi{10.1145/3447245},
  \urlprefix\url{http://dx.doi.org/10.1145/3447245}

\bibitem[{Bosu and Carver(2014)}]{bosu_impact_2014}
Bosu A, Carver JC (2014) Impact of developer reputation on code review outcomes
  in oss projects: an empirical investigation. In: Proceedings of the 8th
  ACM/IEEE International Symposium on Empirical Software Engineering and
  Measurement, ACM, ESEM ’14, p 1–10, \doi{10.1145/2652524.2652544},
  \urlprefix\url{http://dx.doi.org/10.1145/2652524.2652544}

\bibitem[{Breiman(2001)}]{breiman_random_2001}
Breiman L (2001) Random forests. Machine Learning 45(1):5–32,
  \doi{10.1023/a:1010933404324},
  \urlprefix\url{http://dx.doi.org/10.1023/A:1010933404324}

\bibitem[{Bródka et~al.(2012)Bródka, Kazienko, MusiaÅ‚, and
  Skibicki}]{brodka_analysis_2012}
Bródka P, Kazienko P, MusiaÅ‚ K, Skibicki K (2012) Analysis of
  neighbourhoods in multi-layered dynamic social networks. International
  Journal of Computational Intelligence Systems 5(3):582,
  \doi{10.1080/18756891.2012.696922},
  \urlprefix\url{http://dx.doi.org/10.1080/18756891.2012.696922}

\bibitem[{Carillo et~al.(2017)Carillo, Huff, and Chawner}]{carillo_what_2017}
Carillo K, Huff S, Chawner B (2017) What makes a good contributor?
  understanding contributor behavior within large free/open source software
  projects – a socialization perspective. The Journal of Strategic
  Information Systems 26(4):322–359, \doi{10.1016/j.jsis.2017.03.001},
  \urlprefix\url{http://dx.doi.org/10.1016/j.jsis.2017.03.001}

\bibitem[{Casalnuovo et~al.(2015)Casalnuovo, Vasilescu, Devanbu, and
  Filkov}]{casalnuovo_developer_2015}
Casalnuovo C, Vasilescu B, Devanbu P, Filkov V (2015) Developer onboarding in
  {GitHub}: the role of prior social links and language experience. In:
  Proceedings of the 2015 10th Joint Meeting on Foundations of Software
  Engineering, ACM, ESEC/FSE’15, p 817–828, \doi{10.1145/2786805.2786854},
  \urlprefix\url{http://dx.doi.org/10.1145/2786805.2786854}

\bibitem[{Casola et~al.(2009)Casola, Fasolino, Mazzocca, and
  Tramontana}]{casola_ahp-based_2009}
Casola V, Fasolino A, Mazzocca N, Tramontana P (2009) An {AHP}-based framework
  for quality and security evaluation. In: 2009 International Conference on
  Computational Science and Engineering, IEEE, p 405–411,
  \doi{10.1109/cse.2009.391},
  \urlprefix\url{http://dx.doi.org/10.1109/CSE.2009.391}

\bibitem[{Celińska(2018)}]{celinska_coding_2018}
Celińska D (2018) Coding together in a social network: collaboration among
  {GitHub} users. In: Proceedings of the 9th International Conference on Social
  Media and Society, ACM, SMSociety ’18, p 31–40,
  \doi{10.1145/3217804.3217895},
  \urlprefix\url{http://dx.doi.org/10.1145/3217804.3217895}

\bibitem[{Chen et~al.(2012)Chen, L\"{u}, Shang, Zhang, and
  Zhou}]{chen_identifying_2012}
Chen D, L\"{u} L, Shang MS, Zhang YC, Zhou T (2012) Identifying influential
  nodes in complex networks. Physica A: Statistical Mechanics and its
  Applications 391(4):1777–1787, \doi{10.1016/j.physa.2011.09.017},
  \urlprefix\url{http://dx.doi.org/10.1016/j.physa.2011.09.017}

\bibitem[{Chen et~al.(2024)Chen, Kazman, Catolino, Manca, Tamburri, and Van
  Den~Heuvel}]{chen_empirical_2024}
Chen HM, Kazman R, Catolino G, Manca M, Tamburri DA, Van Den~Heuvel WJ (2024)
  An empirical study of social debt in open-source projects: Social drivers and
  the “known devil” community smell. In: Proceedings of the 57th Hawaii
  International Conference on System Sciences,
  \urlprefix\url{https://hdl.handle.net/10125/107255}

\bibitem[{Cheng et~al.(2017)Cheng, Li, Li, Zhao, and
  Liao}]{cheng_developer_2017}
Cheng C, Li B, Li ZY, Zhao YQ, Liao FL (2017) Developer role evolution in open
  source software ecosystem: An explanatory study on {GNOME}. Journal of
  Computer Science and Technology 32(2):396–414,
  \doi{10.1007/s11390-017-1728-9},
  \urlprefix\url{http://dx.doi.org/10.1007/s11390-017-1728-9}

\bibitem[{Chidambaram and Mazrae(2022)}]{chidambaram_bot_2022}
Chidambaram N, Mazrae PR (2022) Bot detection in {GitHub} repositories. In:
  Proceedings of the 19th International Conference on Mining Software
  Repositories, ACM, MSR ’22, \doi{10.1145/3524842.3528520},
  \urlprefix\url{http://dx.doi.org/10.1145/3524842.3528520}

\bibitem[{Cook(2000)}]{cook_detection_2000}
Cook RD (2000) Detection of influential observation in linear regression.
  Technometrics 42(1):65–68, \doi{10.1080/00401706.2000.10485981},
  \urlprefix\url{http://dx.doi.org/10.1080/00401706.2000.10485981}

\bibitem[{Cánovas~Izquierdo and Cabot(2021)}]{canovas_izquierdo_analysis_2021}
Cánovas~Izquierdo JL, Cabot J (2021) On the analysis of non-coding roles in
  open source development: An empirical study of {NPM} package projects.
  Empirical Software Engineering 27(1), \doi{10.1007/s10664-021-10061-x},
  \urlprefix\url{http://dx.doi.org/10.1007/s10664-021-10061-x}

\bibitem[{Decan et~al.(2018)Decan, Mens, and Grosjean}]{decan_empirical_2019}
Decan A, Mens T, Grosjean P (2018) An empirical comparison of dependency
  network evolution in seven software packaging ecosystems. Empirical Software
  Engineering 24(1):381–416, \doi{10.1007/s10664-017-9589-y},
  \urlprefix\url{http://dx.doi.org/10.1007/s10664-017-9589-y}

\bibitem[{Dey and Mockus(2020)}]{dey_effect_2020}
Dey T, Mockus A (2020) Effect of technical and social factors on pull request
  quality for the {NPM} ecosystem. In: Proceedings of the 14th ACM / IEEE
  International Symposium on Empirical Software Engineering and Measurement
  (ESEM), ACM, ESEM ’20, p 1–11, \doi{10.1145/3382494.3410685},
  \urlprefix\url{http://dx.doi.org/10.1145/3382494.3410685}

\bibitem[{Dey et~al.(2020{\natexlab{a}})Dey, Mousavi, Ponce, Fry, Vasilescu,
  Filippova, and Mockus}]{dey_dataset_2020}
Dey T, Mousavi S, Ponce E, Fry T, Vasilescu B, Filippova A, Mockus A
  (2020{\natexlab{a}}) A dataset of bot commits. \doi{10.5281/ZENODO.3694401},
  \urlprefix\url{https://zenodo.org/record/3694401}

\bibitem[{Dey et~al.(2020{\natexlab{b}})Dey, Mousavi, Ponce, Fry, Vasilescu,
  Filippova, and Mockus}]{dey_detecting_2020}
Dey T, Mousavi S, Ponce E, Fry T, Vasilescu B, Filippova A, Mockus A
  (2020{\natexlab{b}}) Detecting and characterizing bots that commit code. In:
  Proceedings of the 17th International Conference on Mining Software
  Repositories, ACM, MSR ’20, p 209–219, \doi{10.1145/3379597.3387478},
  \urlprefix\url{http://dx.doi.org/10.1145/3379597.3387478}

\bibitem[{Dueñas et~al.(2021)Dueñas, Cosentino, Gonzalez-Barahona, del
  Castillo San~Felix, Izquierdo-Cortazar, Cañas-Díaz, and Pérez
  García-Plaza}]{duenas_grimoirelab_2021}
Dueñas S, Cosentino V, Gonzalez-Barahona JM, del Castillo San~Felix A,
  Izquierdo-Cortazar D, Cañas-Díaz L, Pérez García-Plaza A (2021)
  {GrimoireLab}: A toolset for software development analytics. PeerJ Computer
  Science 7:e601, \doi{10.7717/peerj-cs.601},
  \urlprefix\url{http://dx.doi.org/10.7717/peerj-cs.601}

\bibitem[{Fang et~al.(2023)Fang, Herbsleb, and Vasilescu}]{fang_matching_2023}
Fang H, Herbsleb J, Vasilescu B (2023) Matching skills, past collaboration, and
  limited competition: Modeling when open-source projects attract contributors.
  In: Proceedings of the 31st ACM Joint European Software Engineering
  Conference and Symposium on the Foundations of Software Engineering, ACM,
  ESEC/FSE ’23, \doi{10.1145/3611643.3616282},
  \urlprefix\url{http://dx.doi.org/10.1145/3611643.3616282}

\bibitem[{Fershtman and Gandal(2011)}]{fershtman_direct_2011}
Fershtman C, Gandal N (2011) Direct and indirect knowledge spillovers: the
  “social network” of open‐source projects. The RAND Journal of Economics
  42(1):70–91, \doi{10.1111/j.1756-2171.2010.00126.x},
  \urlprefix\url{http://dx.doi.org/10.1111/j.1756-2171.2010.00126.x}

\bibitem[{Forte and Lampe(2013)}]{forte_defining_2013}
Forte A, Lampe C (2013) Defining, understanding, and supporting open
  collaboration: Lessons from the literature. American Behavioral Scientist
  57(5):535–547, \doi{10.1177/0002764212469362},
  \urlprefix\url{http://dx.doi.org/10.1177/0002764212469362}

\bibitem[{Franco-Bedoya et~al.(2017)Franco-Bedoya, Ameller, Costal, and
  Franch}]{franco-bedoya_open_2017}
Franco-Bedoya O, Ameller D, Costal D, Franch X (2017) Open source software
  ecosystems: A systematic mapping. Information and Software Technology
  91:160–185, \doi{10.1016/j.infsof.2017.07.007},
  \urlprefix\url{http://dx.doi.org/10.1016/j.infsof.2017.07.007}

\bibitem[{Geiger et~al.(2021)Geiger, Howard, and Irani}]{geiger_labor_2021}
Geiger RS, Howard D, Irani L (2021) The labor of maintaining and scaling free
  and open-source software projects. Proceedings of the ACM on Human-Computer
  Interaction 5(CSCW1):1–28, \doi{10.1145/3449249},
  \urlprefix\url{http://dx.doi.org/10.1145/3449249}

\bibitem[{Gharehyazie et~al.(2013)Gharehyazie, Posnett, and
  Filkov}]{gharehyazie_social_2013}
Gharehyazie M, Posnett D, Filkov V (2013) Social activities rival patch
  submission for prediction of developer initiation in {OSS} projects. In: 2013
  IEEE International Conference on Software Maintenance, IEEE, p 340–349,
  \doi{10.1109/icsm.2013.45},
  \urlprefix\url{http://dx.doi.org/10.1109/ICSM.2013.45}

\bibitem[{Gharehyazie et~al.(2014)Gharehyazie, Posnett, Vasilescu, and
  Filkov}]{gharehyazie_developer_2015}
Gharehyazie M, Posnett D, Vasilescu B, Filkov V (2014) Developer initiation and
  social interactions in {OSS}: A case study of the {Apache} {Software}
  {Foundation}. Empirical Software Engineering 20(5):1318–1353,
  \doi{10.1007/s10664-014-9332-x},
  \urlprefix\url{http://dx.doi.org/10.1007/s10664-014-9332-x}

\bibitem[{Golzadeh et~al.(2019)Golzadeh, Decan, and
  Mens}]{golzadeh_effect_2019}
Golzadeh M, Decan A, Mens T (2019) On the effect of discussions on pull request
  decisions. In: Proceedings of the 18th Belgium-Netherlands Software Evolution
  Workshop, \urlprefix\url{https://ceur-ws.org/Vol-2605/16.pdf}

\bibitem[{Golzadeh et~al.(2020{\natexlab{a}})Golzadeh, Decan, Legay, and
  Mens}]{golzadeh_ground-truth_2020}
Golzadeh M, Decan A, Legay D, Mens T (2020{\natexlab{a}}) A ground-truth
  dataset to identify bots in {GitHub}. \doi{10.5281/ZENODO.4000388},
  \urlprefix\url{https://zenodo.org/record/4000388}

\bibitem[{Golzadeh et~al.(2020{\natexlab{b}})Golzadeh, Legay, Decan, and
  Mens}]{golzadeh_bot_2020}
Golzadeh M, Legay D, Decan A, Mens T (2020{\natexlab{b}}) Bot or not?:
  Detecting bots in {GitHub} pull request activity based on comment similarity.
  In: Proceedings of the IEEE/ACM 42nd International Conference on Software
  Engineering Workshops, ACM, ICSE ’20, p 31–35,
  \doi{10.1145/3387940.3391503},
  \urlprefix\url{http://dx.doi.org/10.1145/3387940.3391503}

\bibitem[{Golzadeh et~al.(2021)Golzadeh, Decan, Legay, and
  Mens}]{golzadeh_ground-truth_2021}
Golzadeh M, Decan A, Legay D, Mens T (2021) A ground-truth dataset and
  classification model for detecting bots in {GitHub} issue and {PR} comments.
  Journal of Systems and Software 175:110911, \doi{10.1016/j.jss.2021.110911},
  \urlprefix\url{http://dx.doi.org/10.1016/j.jss.2021.110911}

\bibitem[{Gousios et~al.(2014)Gousios, Pinzger, and
  Deursen}]{gousios_exploratory_2014}
Gousios G, Pinzger M, Deursen Av (2014) An exploratory study of the pull-based
  software development model. In: Proceedings of the 36th International
  Conference on Software Engineering, ACM, ICSE ’14,
  \doi{10.1145/2568225.2568260},
  \urlprefix\url{http://dx.doi.org/10.1145/2568225.2568260}

\bibitem[{Gousios et~al.(2015)Gousios, Zaidman, Storey, and
  Deursen}]{gousios_work_2015}
Gousios G, Zaidman A, Storey MA, Deursen Av (2015) Work practices and
  challenges in pull-based development: The integrator’s perspective. In:
  2015 IEEE/ACM 37th IEEE International Conference on Software Engineering,
  IEEE, \doi{10.1109/icse.2015.55},
  \urlprefix\url{http://dx.doi.org/10.1109/ICSE.2015.55}

\bibitem[{Hahn et~al.(2006)Hahn, Moon, and Zhang}]{hahn_impact_2006}
Hahn J, Moon JY, Zhang C (2006) Impact of Social Ties on Open Source Project
  Team Formation, Springer US, p 307–317. \doi{10.1007/0-387-34226-5_31},
  \urlprefix\url{http://dx.doi.org/10.1007/0-387-34226-5_31}

\bibitem[{Hahn et~al.(2008)Hahn, Moon, and Zhang}]{hahn_emergence_2008}
Hahn J, Moon JY, Zhang C (2008) Emergence of new project teams from open source
  software developer networks: Impact of prior collaboration ties. Information
  Systems Research 19(3):369–391, \doi{10.1287/isre.1080.0192},
  \urlprefix\url{http://dx.doi.org/10.1287/isre.1080.0192}

\bibitem[{Hanssen and Dybå(2012)}]{hanssen_theoretical_2012}
Hanssen GK, Dybå T (2012) Theoretical foundations of software ecosystems. In:
  Proceedings of the Fourth International Workshop on Software Ecosystems,
  \urlprefix\url{https://ceur-ws.org/Vol-879/paper1.pdf}

\bibitem[{Holme and Saram\"{a}ki(2019)}]{holme_map_2019}
Holme P, Saram\"{a}ki J (2019) A Map of Approaches to Temporal Networks,
  Springer International Publishing, p 1–24.
  \doi{10.1007/978-3-030-23495-9_1},
  \urlprefix\url{http://dx.doi.org/10.1007/978-3-030-23495-9_1}

\bibitem[{Hou et~al.(2024)Hou, Zhao, Liu, Yang, Wang, Li, Luo, Lo, Grundy, and
  Wang}]{hou_large_2024}
Hou X, Zhao Y, Liu Y, Yang Z, Wang K, Li L, Luo X, Lo D, Grundy J, Wang H
  (2024) Large language models for software engineering: A systematic
  literature review. ACM Transactions on Software Engineering and Methodology
  33(8):1–79, \doi{10.1145/3695988},
  \urlprefix\url{http://dx.doi.org/10.1145/3695988}

\bibitem[{Iyer et~al.(2021)Iyer, Yun, Nagappan, and Hoey}]{iyer_effects_2021}
Iyer RN, Yun SA, Nagappan M, Hoey J (2021) Effects of personality traits on
  pull request acceptance. IEEE Transactions on Software Engineering
  47(11):2632–2643, \doi{10.1109/tse.2019.2960357},
  \urlprefix\url{http://dx.doi.org/10.1109/TSE.2019.2960357}

\bibitem[{Jergensen et~al.(2011)Jergensen, Sarma, and
  Wagstrom}]{jergensen_onion_2011}
Jergensen C, Sarma A, Wagstrom P (2011) The onion patch: migration in open
  source ecosystems. In: Proceedings of the 19th ACM SIGSOFT symposium and the
  13th European conference on Foundations of software engineering, ACM,
  ESEC/FSE’11, \doi{10.1145/2025113.2025127},
  \urlprefix\url{http://dx.doi.org/10.1145/2025113.2025127}

\bibitem[{Katz(2020)}]{katz_librariesio_2020}
Katz J (2020) {Libraries.io} open source repository and dependency metadata.
  \doi{10.5281/ZENODO.3626071},
  \urlprefix\url{https://zenodo.org/record/3626071}

\bibitem[{Khadke et~al.(2012)Khadke, Teh, and Shen}]{khadke_predicting_2012}
Khadke N, Teh MH, Shen M (2012) Predicting acceptance of {GitHub} pull
  requests. Stanford University,
  \urlprefix\url{https://cs229.stanford.edu/proj2012/KhadkeTehShen-PredictingAcceptanceOfGitHubPullRequests.pdf}

\bibitem[{Kivel{\"a} et~al.(2014)Kivel{\"a}, Arenas, Barthelemy, Gleeson,
  Moreno, and Porter}]{kivela_multilayer_2014}
Kivel{\"a} M, Arenas A, Barthelemy M, Gleeson JP, Moreno Y, Porter MA (2014)
  Multilayer networks. Journal of Complex Networks 2(3):203–271,
  \doi{10.1093/comnet/cnu016},
  \urlprefix\url{http://dx.doi.org/10.1093/comnet/cnu016}

\bibitem[{Kononenko et~al.(2018)Kononenko, Rose, Baysal, Godfrey, Theisen, and
  de~Water}]{kononenko_studying_2018}
Kononenko O, Rose T, Baysal O, Godfrey M, Theisen D, de~Water B (2018) Studying
  pull request merges: a case study of {Shopify’s} active merchant. In:
  Proceedings of the 40th International Conference on Software Engineering:
  Software Engineering in Practice, ACM, ICSE ’18,
  \doi{10.1145/3183519.3183542},
  \urlprefix\url{http://dx.doi.org/10.1145/3183519.3183542}

\bibitem[{Kovalenko and Bacchelli(2018)}]{kovalenko_code_2018}
Kovalenko V, Bacchelli A (2018) Code review for newcomers: is it different? In:
  Proceedings of the 11th International Workshop on Cooperative and Human
  Aspects of Software Engineering, ACM, ICSE ’18, p 29–32,
  \doi{10.1145/3195836.3195842},
  \urlprefix\url{http://dx.doi.org/10.1145/3195836.3195842}

\bibitem[{Lee and Carver(2017)}]{lee_are_2017}
Lee A, Carver JC (2017) Are one-time contributors different? a comparison to
  core and periphery developers in {FLOSS} repositories. In: 2017 ACM/IEEE
  International Symposium on Empirical Software Engineering and Measurement
  (ESEM), IEEE, \doi{10.1109/esem.2017.7},
  \urlprefix\url{http://dx.doi.org/10.1109/ESEM.2017.7}

\bibitem[{Legay et~al.(2019)Legay, Decan, and Mens}]{legay_impact_2019}
Legay D, Decan A, Mens T (2019) On the impact of pull request decisions on
  future contributions. In: Belgium-Netherlands Software Evolution Workshop,
  \urlprefix\url{https://ceur-ws.org/Vol-2361/short12.pdf}

\bibitem[{Li et~al.(2022)Li, Yu, Wang, Li, and Wang}]{li_opportunities_2022}
Li Z, Yu Y, Wang T, Li S, Wang H (2022) Opportunities and challenges in
  repeated revisions to pull-requests: An empirical study. Proceedings of the
  ACM on Human-Computer Interaction 6(CSCW2):1–35, \doi{10.1145/3555208},
  \urlprefix\url{http://dx.doi.org/10.1145/3555208}

\bibitem[{Lohr(2021)}]{lohr_sampling_2021}
Lohr SL (2021) Sampling: Design and Analysis. Chapman and Hall/CRC,
  \doi{10.1201/9780429298899},
  \urlprefix\url{http://dx.doi.org/10.1201/9780429298899}

\bibitem[{Lu et~al.(2023)Lu, Yu, Li, Yang, and Zuo}]{lu_llama-reviewer_2023}
Lu J, Yu L, Li X, Yang L, Zuo C (2023) {LLaMA}-{Reviewer}: Advancing code
  review automation with large language models through parameter-efficient
  fine-tuning. In: 2023 IEEE 34th International Symposium on Software
  Reliability Engineering (ISSRE), IEEE, p 647–658,
  \doi{10.1109/issre59848.2023.00026},
  \urlprefix\url{http://dx.doi.org/10.1109/ISSRE59848.2023.00026}

\bibitem[{Ma et~al.(2020)Ma, Chen, Zhang, Feng, Xu, Chen, Zhou, and
  Xu}]{ma_impact_2020}
Ma W, Chen L, Zhang X, Feng Y, Xu Z, Chen Z, Zhou Y, Xu B (2020) Impact
  analysis of cross-project bugs on software ecosystems. In: Proceedings of the
  ACM/IEEE 42nd International Conference on Software Engineering, ACM, ICSE
  ’20, p 100–111, \doi{10.1145/3377811.3380442},
  \urlprefix\url{http://dx.doi.org/10.1145/3377811.3380442}

\bibitem[{Maeprasart et~al.(2023)Maeprasart, Wattanakriengkrai, Kula, Treude,
  and Matsumoto}]{maeprasart_understanding_2023}
Maeprasart V, Wattanakriengkrai S, Kula RG, Treude C, Matsumoto K (2023)
  Understanding the role of external pull requests in the {NPM} ecosystem.
  Empirical Software Engineering 28(4), \doi{10.1007/s10664-023-10315-w},
  \urlprefix\url{http://dx.doi.org/10.1007/s10664-023-10315-w}

\bibitem[{Mayring(2014)}]{mayring_qualitative_2014}
Mayring P (2014) Qualitative content analysis: theoretical foundation, basic
  procedures and software solution. AUT

\bibitem[{Meijer(2023)}]{meijer_influence_2023}
Meijer W (2023) {The} {influence} {of} {ecosystem-wide} {experience} {and}
  {collaboration} {on} {pull} {request} {acceptance} {in} {open-source}
  {software} {ecosystems}. Master's thesis, University of Groningen,
  \urlprefix\url{https://fse.studenttheses.ub.rug.nl/31331/}

\bibitem[{Meijer et~al.(2024)Meijer, Riveni, and Rastogi}]{meijer_2024_dataset}
Meijer W, Riveni M, Rastogi A (2024) Replication package (obfuscated):
  Ecosystem-wide influences on pull request decisions.
  \urlprefix\url{https://tinyurl.com/5449javb}

\bibitem[{Meneely and Williams(2011)}]{meneely_socio-technical_2011}
Meneely A, Williams L (2011) Socio-technical developer networks: should we
  trust our measurements? In: Proceedings of the 33rd International Conference
  on Software Engineering, ACM, ICSE11, \doi{10.1145/1985793.1985832},
  \urlprefix\url{http://dx.doi.org/10.1145/1985793.1985832}

\bibitem[{Méndez-Durón and García(2008)}]{mendez-duron_returns_2009}
Méndez-Durón R, García CE (2008) Returns from social capital in open source
  software networks. Journal of Evolutionary Economics 19(2):277–295,
  \doi{10.1007/s00191-008-0125-5},
  \urlprefix\url{http://dx.doi.org/10.1007/s00191-008-0125-5}

\bibitem[{Müller and Rosenkranz(2023)}]{muller_role_2023}
Müller M, Rosenkranz C (2023) The role of dependency networks in developer
  participation decisions in open source software ecosystems: An application of
  stochastic actor-oriented models. In: Proceedings of the 56th Hawaii
  International Conference on System Sciences, pp 689--698,
  \urlprefix\url{https://hdl.handle.net/10125/102715}

\bibitem[{Newman(2018)}]{newman_measures_2018}
Newman M (2018) Measures and metrics. Oxford University Press,
  \doi{10.1093/oso/9780198805090.003.0007},
  \urlprefix\url{http://dx.doi.org/10.1093/oso/9780198805090.003.0007}

\bibitem[{Palyart et~al.(2018)Palyart, Murphy, and
  Masrani}]{palyart_study_2018}
Palyart M, Murphy GC, Masrani V (2018) A study of social interactions in open
  source component use. IEEE Transactions on Software Engineering
  44(12):1132–1145, \doi{10.1109/tse.2017.2756043},
  \urlprefix\url{http://dx.doi.org/10.1109/TSE.2017.2756043}

\bibitem[{Panichella et~al.(2014)Panichella, Bavota, Penta, Canfora, and
  Antoniol}]{panichella_how_2014}
Panichella S, Bavota G, Penta MD, Canfora G, Antoniol G (2014) How
  developers’ collaborations identified from different sources tell us about
  code changes. In: 2014 IEEE International Conference on Software Maintenance
  and Evolution, IEEE, p 251–260, \doi{10.1109/icsme.2014.47},
  \urlprefix\url{http://dx.doi.org/10.1109/ICSME.2014.47}

\bibitem[{Pedregosa et~al.(2011)Pedregosa, Varoquaux, Gramfort, Michel,
  Thirion, Grisel, Blondel, Prettenhofer, Weiss, Dubourg, Vanderplas, Passos,
  Cournapeau, Brucher, Perrot, and Duchesnay}]{pedregosa_scikit-learn_2011}
Pedregosa F, Varoquaux G, Gramfort A, Michel V, Thirion B, Grisel O, Blondel M,
  Prettenhofer P, Weiss R, Dubourg V, Vanderplas J, Passos A, Cournapeau D,
  Brucher M, Perrot M, Duchesnay {\'E} (2011) {Scikit-learn}: Machine learning
  in {Python}. Journal of Machine Learning Research 12(85):2825--2830,
  \urlprefix\url{http://jmlr.org/papers/v12/pedregosa11a.html}

\bibitem[{Peng et~al.(2013)Peng, Wan, and Woodlock}]{peng_network_2013}
Peng G, Wan Y, Woodlock P (2013) Network ties and the success of open source
  software development. The Journal of Strategic Information Systems
  22(4):269–281, \doi{10.1016/j.jsis.2013.05.001},
  \urlprefix\url{http://dx.doi.org/10.1016/j.jsis.2013.05.001}

\bibitem[{P{\'e}rez-Verdejo et~al.(2021)P{\'e}rez-Verdejo,
  S{\'a}nchez-Garc{\'i}a, Ochar{\'a}n-Hern{\'a}ndez, Mezura-Montes, and
  Cort{\'e}s-Verdín}]{perez-verdejo_requirements_2021}
P{\'e}rez-Verdejo JM, S{\'a}nchez-Garc{\'i}a {\'A}J, Ochar{\'a}n-Hern{\'a}ndez
  JO, Mezura-Montes E, Cort{\'e}s-Verdín K (2021) Requirements and {GitHub}
  issues: An automated approach for quality requirements classification.
  Programming and Computer Software 47(8):704–721,
  \doi{10.1134/s0361768821080193},
  \urlprefix\url{http://dx.doi.org/10.1134/S0361768821080193}

\bibitem[{Pinto et~al.(2018)Pinto, Dias, and Steinmacher}]{pinto_who_2018}
Pinto G, Dias LF, Steinmacher I (2018) Who gets a patch accepted first?:
  comparing the contributions of employees and volunteers. In: Proceedings of
  the 11th International Workshop on Cooperative and Human Aspects of Software
  Engineering, ACM, ICSE ’18, \doi{10.1145/3195836.3195858},
  \urlprefix\url{http://dx.doi.org/10.1145/3195836.3195858}

\bibitem[{Qiu et~al.(2019)Qiu, Nolte, Brown, Serebrenik, and
  Vasilescu}]{qiu_going_2019}
Qiu HS, Nolte A, Brown A, Serebrenik A, Vasilescu B (2019) Going farther
  together: The impact of social capital on sustained participation in open
  source. In: 2019 IEEE/ACM 41st International Conference on Software
  Engineering (ICSE), IEEE, p 688–699, \doi{10.1109/icse.2019.00078},
  \urlprefix\url{http://dx.doi.org/10.1109/ICSE.2019.00078}

\bibitem[{Rastogi(2016)}]{rastogi_biases_2016}
Rastogi A (2016) Do biases related to geographical location influence
  work-related decisions in {GitHub}? In: Proceedings of the 38th International
  Conference on Software Engineering Companion, ACM, ICSE ’16, p 665–667,
  \doi{10.1145/2889160.2891035},
  \urlprefix\url{http://dx.doi.org/10.1145/2889160.2891035}

\bibitem[{Rastogi and Gousios(2021)}]{rastogi_how_2021}
Rastogi A, Gousios G (2021) How does software change?
  \doi{10.48550/ARXIV.2106.01885},
  \urlprefix\url{https://arxiv.org/abs/2106.01885}

\bibitem[{Rastogi et~al.(2015)Rastogi, Thummalapenta, Zimmermann, Nagappan, and
  Czerwonka}]{rastogi_ramp-up_2015}
Rastogi A, Thummalapenta S, Zimmermann T, Nagappan N, Czerwonka J (2015)
  Ramp-up journey of new hires: Tug of war of aids and impediments. In: 2015
  ACM/IEEE International Symposium on Empirical Software Engineering and
  Measurement (ESEM), IEEE, p 1–10, \doi{10.1109/esem.2015.7321212},
  \urlprefix\url{http://dx.doi.org/10.1109/ESEM.2015.7321212}

\bibitem[{Rastogi et~al.(2018)Rastogi, Nagappan, Gousios, and van~der
  Hoek}]{rastogi_relationship_2018}
Rastogi A, Nagappan N, Gousios G, van~der Hoek A (2018) Relationship between
  geographical location and evaluation of developer contributions in {GitHub}.
  In: Proceedings of the 12th ACM/IEEE International Symposium on Empirical
  Software Engineering and Measurement, ACM, ESEM ’18,
  \doi{10.1145/3239235.3240504},
  \urlprefix\url{http://dx.doi.org/10.1145/3239235.3240504}

\bibitem[{Rehman et~al.(2022)Rehman, Wang, Kula, Ishio, and
  Matsumoto}]{rehman_newcomer_2022}
Rehman I, Wang D, Kula RG, Ishio T, Matsumoto K (2022) Newcomer
  {OSS}-candidates: Characterizing contributions of novice developers to
  {GitHub}. Empirical Software Engineering 27(5),
  \doi{10.1007/s10664-022-10163-0},
  \urlprefix\url{http://dx.doi.org/10.1007/s10664-022-10163-0}

\bibitem[{Rombaut et~al.(2023)Rombaut, Cogo, Adams, and
  Hassan}]{rombaut_theres_2023}
Rombaut B, Cogo FR, Adams B, Hassan AE (2023) There’s no such thing as a free
  lunch: Lessons learned from exploring the overhead introduced by the
  {Greenkeeper} dependency bot in {NPM}. ACM Transactions on Software
  Engineering and Methodology 32(1):1–40, \doi{10.1145/3522587},
  \urlprefix\url{http://dx.doi.org/10.1145/3522587}

\bibitem[{Schreiber and Zylka(2020)}]{schreiber_social_2020}
Schreiber RR, Zylka MP (2020) Social network analysis in software development
  projects: A systematic literature review. International Journal of Software
  Engineering and Knowledge Engineering 30(03):321–362,
  \doi{10.1142/s021819402050014x},
  \urlprefix\url{http://dx.doi.org/10.1142/S021819402050014X}

\bibitem[{Seaman(1999)}]{seaman_qualitative_1999}
Seaman C (1999) Qualitative methods in empirical studies of software
  engineering. IEEE Transactions on Software Engineering 25(4):557–572,
  \doi{10.1109/32.799955}, \urlprefix\url{http://dx.doi.org/10.1109/32.799955}

\bibitem[{Shah(2006)}]{shah_motivation_2006}
Shah SK (2006) Motivation, governance, and the viability of hybrid forms in
  open source software development. Management Science 52(7):1000–1014,
  \doi{10.1287/mnsc.1060.0553},
  \urlprefix\url{http://dx.doi.org/10.1287/mnsc.1060.0553}

\bibitem[{Soares et~al.(2015{\natexlab{a}})Soares, De~Lima~Junior, Murta, and
  Plastino}]{soares_rejection_2015}
Soares DM, De~Lima~Junior ML, Murta L, Plastino A (2015{\natexlab{a}})
  Rejection factors of pull requests filed by core team developers in software
  projects with high acceptance rates. In: 2015 IEEE 14th International
  Conference on Machine Learning and Applications (ICMLA), IEEE, p 960–965,
  \doi{10.1109/icmla.2015.41},
  \urlprefix\url{http://dx.doi.org/10.1109/ICMLA.2015.41}

\bibitem[{Soares et~al.(2015{\natexlab{b}})Soares, de~Lima~Júnior, Murta, and
  Plastino}]{soares_acceptance_2015}
Soares DM, de~Lima~Júnior ML, Murta L, Plastino A (2015{\natexlab{b}})
  Acceptance factors of pull requests in open-source projects. In: Proceedings
  of the 30th Annual ACM Symposium on Applied Computing, ACM, SAC 2015,
  \doi{10.1145/2695664.2695856},
  \urlprefix\url{http://dx.doi.org/10.1145/2695664.2695856}

\bibitem[{Soliman et~al.(2021)Soliman, Galster, and
  Avgeriou}]{soliman_exploratory_2021}
Soliman M, Galster M, Avgeriou P (2021) An Exploratory Study on Architectural
  Knowledge in Issue Tracking Systems, Springer International Publishing, p
  117–133. \doi{10.1007/978-3-030-86044-8_8},
  \urlprefix\url{http://dx.doi.org/10.1007/978-3-030-86044-8_8}

\bibitem[{Soto et~al.(2017)Soto, Coker, and Le~Goues}]{soto_analyzing_2017}
Soto M, Coker Z, Le~Goues C (2017) Analyzing the impact of social attributes on
  commit integration success. In: 2017 IEEE/ACM 14th International Conference
  on Mining Software Repositories (MSR), IEEE, p 483–486,
  \doi{10.1109/msr.2017.34},
  \urlprefix\url{http://dx.doi.org/10.1109/MSR.2017.34}

\bibitem[{de~Souza et~al.(2016)de~Souza, Figueira~Filho, Miranda, Ferreira,
  Treude, and Singer}]{de_souza_social_2016}
de~Souza CR, Figueira~Filho F, Miranda M, Ferreira RP, Treude C, Singer L
  (2016) The social side of software platform ecosystems. In: Proceedings of
  the 2016 CHI Conference on Human Factors in Computing Systems, ACM, CHI’16,
  p 3204–3214, \doi{10.1145/2858036.2858431},
  \urlprefix\url{http://dx.doi.org/10.1145/2858036.2858431}

\bibitem[{Steinmacher et~al.(2015)Steinmacher, Conte, Gerosa, and
  Redmiles}]{steinmacher_social_2015}
Steinmacher I, Conte T, Gerosa MA, Redmiles D (2015) Social barriers faced by
  newcomers placing their first contribution in open source software projects.
  In: Proceedings of the 18th ACM Conference on Computer Supported Cooperative
  Work \& Social Computing, ACM, CSCW ’15, p 1379–1392,
  \doi{10.1145/2675133.2675215},
  \urlprefix\url{http://dx.doi.org/10.1145/2675133.2675215}

\bibitem[{Steinmacher et~al.(2018{\natexlab{a}})Steinmacher, Gerosa, Conte, and
  Redmiles}]{steinmacher_overcoming_2019}
Steinmacher I, Gerosa M, Conte TU, Redmiles DF (2018{\natexlab{a}}) Overcoming
  social barriers when contributing to open source software projects. Computer
  Supported Cooperative Work (CSCW) 28(1–2):247–290,
  \doi{10.1007/s10606-018-9335-z},
  \urlprefix\url{http://dx.doi.org/10.1007/s10606-018-9335-z}

\bibitem[{Steinmacher et~al.(2018{\natexlab{b}})Steinmacher, Pinto, Wiese, and
  Gerosa}]{steinmacher_almost_2018}
Steinmacher I, Pinto G, Wiese IS, Gerosa MA (2018{\natexlab{b}}) Almost there:
  a study on quasi-contributors in open source software projects. In:
  Proceedings of the 40th International Conference on Software Engineering,
  ACM, ICSE ’18, p 256–266, \doi{10.1145/3180155.3180208},
  \urlprefix\url{http://dx.doi.org/10.1145/3180155.3180208}

\bibitem[{Steinmacher et~al.(2019)Steinmacher, Treude, and
  Gerosa}]{steinmacher_let_2019}
Steinmacher I, Treude C, Gerosa MA (2019) Let me in: Guidelines for the
  successful onboarding of newcomers to open source projects. IEEE Software
  36(4):41–49, \doi{10.1109/ms.2018.110162131},
  \urlprefix\url{http://dx.doi.org/10.1109/MS.2018.110162131}

\bibitem[{Stickgold et~al.(2013)Stickgold, Lofdahl, and
  Farry}]{stickgold_trust_2013}
Stickgold E, Lofdahl C, Farry M (2013) Trust Metrics and Results for Social
  Media Analysis, Springer Berlin Heidelberg, p 458–465.
  \doi{10.1007/978-3-642-37210-0_50},
  \urlprefix\url{http://dx.doi.org/10.1007/978-3-642-37210-0_50}

\bibitem[{Subramanian et~al.(2022)Subramanian, Rehman, Nagappan, and
  Kula}]{subramanian_analyzing_2022}
Subramanian VN, Rehman I, Nagappan M, Kula RG (2022) Analyzing first
  contributions on {GitHub}: What do newcomers do? IEEE Software
  39(1):93–101, \doi{10.1109/ms.2020.3041241},
  \urlprefix\url{http://dx.doi.org/10.1109/MS.2020.3041241}

\bibitem[{Syeed et~al.(2014)Syeed, Hansen, Hammouda, and
  Manikas}]{syeed_socio-technical_2014}
Syeed MMM, Hansen KM, Hammouda I, Manikas K (2014) Socio-technical congruence
  in the {Ruby} ecosystem. In: Proceedings of The International Symposium on
  Open Collaboration, ACM, OpenSym ’14, p 1–9,
  \doi{10.1145/2641580.2641586},
  \urlprefix\url{http://dx.doi.org/10.1145/2641580.2641586}

\bibitem[{Terrell et~al.(2017)Terrell, Kofink, Middleton, Rainear, Murphy-Hill,
  Parnin, and Stallings}]{terrell_gender_2017}
Terrell J, Kofink A, Middleton J, Rainear C, Murphy-Hill E, Parnin C, Stallings
  J (2017) Gender differences and bias in open source: pull request acceptance
  of women versus men. PeerJ Computer Science 3:e111,
  \doi{10.7717/peerj-cs.111},
  \urlprefix\url{http://dx.doi.org/10.7717/peerj-cs.111}

\bibitem[{Trinkenreich et~al.(2020)Trinkenreich, Guizani, Wiese, Sarma, and
  Steinmacher}]{trinkenreich_hidden_2020}
Trinkenreich B, Guizani M, Wiese I, Sarma A, Steinmacher I (2020) Hidden
  figures: Roles and pathways of successful {OSS} contributors. Proceedings of
  the ACM on Human-Computer Interaction 4(CSCW2):1–22, \doi{10.1145/3415251},
  \urlprefix\url{http://dx.doi.org/10.1145/3415251}

\bibitem[{Trinkenreich et~al.(2022)Trinkenreich, Guizani, Wiese, Conte, Gerosa,
  Sarma, and Steinmacher}]{trinkenreich_pots_2022}
Trinkenreich B, Guizani M, Wiese I, Conte T, Gerosa M, Sarma A, Steinmacher I
  (2022) Pots of gold at the end of the rainbow: What is success for open
  source contributors? IEEE Transactions on Software Engineering
  48(10):3940–3953, \doi{10.1109/tse.2021.3108032},
  \urlprefix\url{http://dx.doi.org/10.1109/TSE.2021.3108032}

\bibitem[{Tsay et~al.(2014)Tsay, Dabbish, and Herbsleb}]{tsay_influence_2014}
Tsay J, Dabbish L, Herbsleb J (2014) Influence of social and technical factors
  for evaluating contribution in {GitHub}. In: Proceedings of the 36th
  International Conference on Software Engineering, ACM, ICSE ’14, p
  356–366, \doi{10.1145/2568225.2568315},
  \urlprefix\url{http://dx.doi.org/10.1145/2568225.2568315}

\bibitem[{Valiev et~al.(2018)Valiev, Vasilescu, and
  Herbsleb}]{valiev_ecosystem-level_2018}
Valiev M, Vasilescu B, Herbsleb J (2018) Ecosystem-level determinants of
  sustained activity in open-source projects: a case study of the {PyPI}
  ecosystem. In: Proceedings of the 2018 26th ACM Joint Meeting on European
  Software Engineering Conference and Symposium on the Foundations of Software
  Engineering, ACM, ESEC/FSE ’18, \doi{10.1145/3236024.3236062},
  \urlprefix\url{http://dx.doi.org/10.1145/3236024.3236062}

\bibitem[{Vasilescu et~al.(2015)Vasilescu, Serebrenik, and
  Filkov}]{vasilescu_data_2015}
Vasilescu B, Serebrenik A, Filkov V (2015) A data set for social diversity
  studies of {GitHub} teams. In: 2015 IEEE/ACM 12th Working Conference on
  Mining Software Repositories, IEEE, p 514–517, \doi{10.1109/msr.2015.77},
  \urlprefix\url{http://dx.doi.org/10.1109/MSR.2015.77}

\bibitem[{Wang(2012)}]{wang_survival_2012}
Wang J (2012) Survival factors for free open source software projects: A
  multi-stage perspective. European Management Journal 30(4):352–371,
  \doi{10.1016/j.emj.2012.03.001},
  \urlprefix\url{http://dx.doi.org/10.1016/j.emj.2012.03.001}

\bibitem[{Wattanakriengkrai et~al.(2023)Wattanakriengkrai, Wang, Kula, Treude,
  Thongtanunam, Ishio, and Matsumoto}]{wattanakriengkrai_giving_2023}
Wattanakriengkrai S, Wang D, Kula RG, Treude C, Thongtanunam P, Ishio T,
  Matsumoto K (2023) Giving back: Contributions congruent to library dependency
  changes in a software ecosystem. IEEE Transactions on Software Engineering
  49(4):2566–2579, \doi{10.1109/tse.2022.3225197},
  \urlprefix\url{http://dx.doi.org/10.1109/TSE.2022.3225197}

\bibitem[{Wermke et~al.(2022)Wermke, W\"{o}hler, Klemmer, Fourné, Acar, and
  Fahl}]{wermke_committed_2022}
Wermke D, W\"{o}hler N, Klemmer JH, Fourné M, Acar Y, Fahl S (2022) Committed
  to trust: A qualitative study on security \& trust in open source software
  projects. In: 2022 IEEE Symposium on Security and Privacy (SP), IEEE, p
  1880–1896, \doi{10.1109/sp46214.2022.9833686},
  \urlprefix\url{http://dx.doi.org/10.1109/SP46214.2022.9833686}

\bibitem[{Wiese et~al.(2016)Wiese, Da~Silva, Steinmacher, Treude, and
  Gerosa}]{wiese_who_2016}
Wiese IS, Da~Silva JT, Steinmacher I, Treude C, Gerosa MA (2016) Who is who in
  the mailing list? comparing six disambiguation heuristics to identify
  multiple addresses of a participant. In: 2016 IEEE International Conference
  on Software Maintenance and Evolution (ICSME), IEEE, p 345–355,
  \doi{10.1109/icsme.2016.13},
  \urlprefix\url{http://dx.doi.org/10.1109/ICSME.2016.13}

\bibitem[{Xiao et~al.(2024)Xiao, Hata, Treude, and
  Matsumoto}]{xiao_generative_2024}
Xiao T, Hata H, Treude C, Matsumoto K (2024) Generative {AI} for pull request
  descriptions: Adoption, impact, and developer interventions. Proceedings of
  the ACM on Software Engineering 1(FSE):1043–1065, \doi{10.1145/3643773},
  \urlprefix\url{http://dx.doi.org/10.1145/3643773}

\bibitem[{Yu et~al.(2016)Yu, Yin, Wang, Yang, and Wang}]{yu_determinants_2016}
Yu Y, Yin G, Wang T, Yang C, Wang H (2016) Determinants of pull-based
  development in the context of continuous integration. Science China
  Information Sciences 59(8), \doi{10.1007/s11432-016-5595-8},
  \urlprefix\url{http://dx.doi.org/10.1007/s11432-016-5595-8}

\bibitem[{Zampetti et~al.(2019)Zampetti, Bavota, Canfora, and
  Penta}]{zampetti_study_2019}
Zampetti F, Bavota G, Canfora G, Penta MD (2019) A study on the interplay
  between pull request review and continuous integration builds. In: 2019 IEEE
  26th International Conference on Software Analysis, Evolution and
  Reengineering (SANER), IEEE, p 38–48, \doi{10.1109/saner.2019.8667996},
  \urlprefix\url{http://dx.doi.org/10.1109/SANER.2019.8667996}

\bibitem[{Zhang et~al.(2022)Zhang, Yu, Wang, Rastogi, and
  Wang}]{zhang_pull_2022-1}
Zhang X, Yu Y, Wang T, Rastogi A, Wang H (2022) Pull request latency explained:
  an empirical overview. Empirical Software Engineering 27(6),
  \doi{10.1007/s10664-022-10143-4},
  \urlprefix\url{http://dx.doi.org/10.1007/s10664-022-10143-4}

\bibitem[{Zhang et~al.(2023)Zhang, Yu, Gousios, and Rastogi}]{zhang_pull_2022}
Zhang X, Yu Y, Gousios G, Rastogi A (2023) Pull request decisions explained: An
  empirical overview. IEEE Transactions on Software Engineering
  49(2):849–871, \doi{10.1109/tse.2022.3165056},
  \urlprefix\url{http://dx.doi.org/10.1109/TSE.2022.3165056}

\bibitem[{Z\"{o}ller et~al.(2020)Z\"{o}ller, Morgan, and
  Schr\"{o}der}]{zoller_topology_2020}
Z\"{o}ller N, Morgan JH, Schr\"{o}der T (2020) A topology of groups: What
  {GitHub} can tell us about online collaboration. Technological Forecasting
  and Social Change 161:120291, \doi{10.1016/j.techfore.2020.120291},
  \urlprefix\url{http://dx.doi.org/10.1016/j.techfore.2020.120291}

\end{thebibliography}
% \fi

    \appendix

    % \clearpage
    
\revOneAdd{

\section{Calculating Collaboration Metrics}
\label{app:so-degree}
    
    This appendix provides a formal definition of the metrics used to calculate second-order degree centrality (Equation~\ref{eq:ml-so-degree}) and link strength (Equation~\ref{eq:link-strength}), which we used to measure \qt{ecosystem-wide community standing} and \qt{direct collaboration,} respectively.
    We apply these metrics to multi-layered graphs using weighted layers \citep{kivela_multilayer_2014}.
    Our metrics explicitly account for time, as not doing so would inflate them \citep{holme_map_2019}. \\
    
    We can define an undirected multi-layer graph as $\Gamma = (V, L, E)$ where $V$ is a set of vertices (i.e., users), $L$ a set of layers (i.e., collaboration types), and $E = \{E_\lambda ~|~ \lambda \in L\}$ the \textit{set of sets of edges} of the different layers, such that the set $E_\lambda$ contains all edges of layer $\lambda$ and each edge $\langle u, v, p, t \rangle_\lambda \in E_\lambda$ is a quadruple containing the connected users $u, v$, the project $p$ in which they collaborated, and a timestamp $t$ indicating when they collaborated.
    We use $E_\lambda(u)$ as shorthand for all $\lambda$-edges connected to vertex $u$.
    The following sections provide a top-down explanation of our metrics.

\subsection{Second-order Degree Centrality}
    
    We define second-order degree centrality $C$ as:
    
    \begin{equation}\label{eq:ml-so-degree}
        C(u, p, t) = \sum^{L^2}_{\lambda, \mu} w_\lambda \cdot w_\mu \cdot d(u,p,t,\lambda, \mu)
    \end{equation}
    
    \noindent such that we calculate the centrality for user $u$ in project $p$ at time $t$.
    This calculates the weighted sum of \textit{single-layer} second-order degree $d$ with respect to all pairs of layers $\lambda, \mu \in L$ using weights $w_\lambda, w_\mu$.
    We defer an explanation for weights until later in Equation~\ref{eq:layer-weights}.\\
    
    We define $d$ as follows:
    
    \begin{equation}\label{eq:so-degree}
    d(u,p,t,\lambda, \mu) = \begin{cases}
        0 & \text{if}~ N_\lambda (u, p, t) = \emptyset,\\
         \displaystyle\frac{1}{|N_\lambda (u, p, t)|} \sum^{N_\lambda (u, p, t)}_{\langle v, t'\rangle_{\lambda}} |N'_\mu (v, t')| & \text{otherwise}
    \end{cases}
        % d(u,p,t,\lambda, \mu) = \frac{1}{|N_\lambda (u, p, t)|} \sum^{N_\lambda (u, p, t)}_{\langle v, t'\rangle_{\lambda}} |N'_\mu (v, t')| ~\text{or}~ 0 ~\text{if}~ N_\lambda (u, p, t) = \emptyset
    \end{equation}
    
    \noindent where $N_\lambda(u, p, t)$ is the set of $\lambda$-edges connecting $u$ to its neighbors $v$, and $N_\mu'(v, t')$ is the set of $\mu$-edges connected to $v$.
    If $u$ has no $\lambda$-edges, the metric is $0$.
    Note that, although we defined edges as a quadruple, we omit the project and source vertex in the summation because they are irrelevant in Equation~\ref{eq:so-degree}.\\
    
    We define $N_\lambda$ as follows:
    
    \begin{equation}\label{eq:fo-neighbors}
        N_\lambda (u, p, t) = \{ \langle v, t' \rangle_\lambda ~|~ \langle u, v, q, t' \rangle_\lambda \in E_\lambda(u) \land t' < t \land v \not = u \land q \not = p\}
    \end{equation}
    
    \noindent which yields all $\lambda$-edges connected to $u$, for which the time of creation $t'$ proceeds the time at which the metric is calculated $t$.
    In other words, we do not consider edges (i.e., collaborations) from the future.
    Further, we exclude any self-loops as these indicate a collaboration between a user and themselves (which, by definition, is not collaboration), and we only look at collaborations outside the focal project $p$ (because we are interested in ecosystem collaborations). \\
    
    We define $N_\lambda'$ as follows:
    
    \begin{equation}\label{eq:so-neighbors}
        N'_\lambda(u, t) = \{ \langle u, v, q, t' \rangle_\lambda ~|~ \langle u, v, q, t' \rangle_\lambda \in E_\lambda(u) \land t' < t \land v \not = u \}
    \end{equation}
    
    \noindent which is similar to $N_\lambda$, however, it does not enforce collaborations in ecosystem projects.
    We can relieve this constraint because we are interested in the second-order neighbors' \textit{complete} ecosystem participation.
    Rather, we exclude it in $N_\lambda$ because first-order intra-project collaborations would confound our results to intra-project factors, which are known to relate to pull request acceptance \citep{gharehyazie_social_2013, gharehyazie_developer_2015, carillo_what_2017, bosu_impact_2014}.
    This is not the case for $N'_\lambda$ because we excluded self-loops in $N_\lambda$. \\
    
    Consequently, we have a metric that calculates the multi-layer second-order degree centrality.
    Going back to Equation~\ref{eq:ml-so-degree}, the reason we calculate the weighted sum of \textit{all pairs} of layers $\lambda, \mu \in L^2$ is that there might exist cases where $u$ and $v$ are connected through a $\lambda$ edge, however, $v$ has no further $\lambda$-edges, whereas it could have many $\mu$-edges.
    Not accounting for both edge types would thus miss relevant connections, deflating the metric.\\
    
    Finally, we weigh each pair of layers' contribution using:
    
    \begin{equation}\label{eq:layer-weights}
        w_\lambda = 1 - \frac{|E_\lambda|}{\sum^L_\mu|E_\mu|}
    \end{equation}
    
    \noindent which weighs each layer inverse-proportionally to the number of edges in that layer because some layers have substantially more edges than others.
    For example, there are about $70$ times more edges due to two users participating in the same issue discussion compared to one user integrating the pull request of another.
    If unaddressed, that layer would be $70$ times more impactful on the metric.
    Because these values are small and unintuitive, we normalize them such that $\sum^L_\lambda w_\lambda = 1$.

\subsection{Link Strength}

    Using some of the definitions introduced for second-order degree centrality, we can calculate link strength $S$ as follows:
    
    \begin{equation}\label{eq:link-strength}
        S(u, v, p, t) = \sum^L_\lambda w_\lambda \cdot |e_\lambda(u, v, p, t)|
    \end{equation}
    
    \noindent taking the weighted sum of all $\lambda$-edge counts connecting $u$ and $v$ at time $t$:
    
    \begin{equation}
        e_\lambda(u, v, p, t) = \{\langle u, v, q,t'\rangle_\lambda ~|~ \langle u, v, q, t'\rangle \in E_\lambda \land t' < t \land q \not = p \}
    \end{equation}
    
    \noindent which is the set of edges connecting $u$ and $v$ at times $t'$ created before $t$ in an ecosystem project.
    
}

\end{document}